\documentclass[12pt]{article}
\pdfoutput=1

\usepackage{amsmath,amsfonts,amssymb,amsthm,amssymb,bbm,bm}
\usepackage{pdfsync}
\usepackage{graphicx,import}
\usepackage[latin1]{inputenc}
\usepackage{hyperref}
\usepackage{empheq}
\usepackage[all]{xy}
\usepackage{stmaryrd}
\usepackage{rotating}
\usepackage{color}  
\usepackage{slashed}
\usepackage{cancel}
\usepackage{array, makecell} %
\usepackage{comment}
\usepackage{tikz}
\usepackage{hhline}
\usepackage{comment}

\usepackage{amsthm}
\usepackage[lastexercise,answerdelayed]{exercise}
\usepackage{multicol}
\newtheorem{theorem}{Theorem}[section]
\newtheorem{exercise}[theorem]{Exercise}

\newcommand{\p}{\partial}

\newcommand{\bit}{\begin{itemize}}
\newcommand{\eit}{\end{itemize}}
\newcommand{\bd}{\begin{description}}
\newcommand{\ed}{\end{description}}

\newcommand{\bc}{\begin{center}}
\newcommand{\ec}{\end{center}}
\newcommand{\be}{\begin{equation}}
\newcommand{\ee}{\end{equation}}
\newcommand{\bea}{\begin{eqnarray}}
\newcommand{\eea}{\end{eqnarray}}
\newcommand{\bs}{\begin{subequations}}
\newcommand{\es}{\end{subequations}}

\newcommand{\ve}{\varepsilon}

\def\p{\partial}

\def\CJ{\mathcal{J}}

\def\bz{\bar{z} }
\def\bh{\bar{h} }
\def\bw{\bar{w} }

\catcode`,\active

\catcode`\,12

\usepackage{cite}
\numberwithin{equation}{section}

\begin{document}
\begin{titlepage}
\unitlength = 1mm
\ \\
\vskip 2cm
\begin{center}
{\LARGE{\textsc{Lectures on Celestial Holography}}}\\
\vspace{0.8cm}
Ana-Maria Raclariu
\vspace{0.4cm}
\smallskip \\ \textit{Perimeter Institute for Theoretical Physics, Waterloo, ON N2L 2Y5, Canada}
\vspace{0.4cm}\\
\small{ {araclariu@perimeterinstitute.ca}} 
\vspace{2cm}

\begin{abstract}
These notes consist of 3 lectures on celestial holography given at the Pre-Strings school 2021. We start by reviewing how semiclassically, the subleading soft graviton theorem implies an enhancement of the Lorentz symmetry of scattering in four-dimensional asymptotically flat gravity to Virasoro. This leads to the construction of celestial amplitudes as $\mathcal{S}$-matrices computed in a basis of boost eigenstates. Both massless and massive asymptotic states are recast as insertions on the celestial sphere transforming as global conformal primaries under the Lorentz SL$(2, \mathbb{C})$. We conclude with an overview of celestial symmetries and the constraints they impose on celestial scattering.
\end{abstract}

\vspace{0.5cm}
\end{center}
\vspace{0.8cm}

\end{titlepage}

%\end{titlepage}S3

\tableofcontents

\section{Introduction}
AdS/CFT \cite{Maldacena:1997re, Aharony:1999ti, Ryu:2006bv} provided a concrete realization of the holographic principle \cite{Susskind:1994vu, Bousso:2002ju}: a theory of gravity in an arbitrary number of dimensions should be dual to a quantum theory in one dimension less. A concrete realization of this duality in any but asymptotically negatively curved backgrounds remains an important open problem.  The goal of these lectures is to review some of the recent developments addressing this problem in asymptotically flat spacetimes (AFS). 

In the past decade we learned that gravity and gauge theory in AFS are governed in the infrared  by a triangular equivalence: soft theorems can be recast as conservation laws associated with large gauge symmetries \cite{He:2014laa,He:2014cra}, while memory effects are an observable signature thereof \cite{Strominger:2014pwa, Pasterski:2015zua}. (See \cite{Strominger:2017zoo} for a detailed review.) These developments led to the proposal that gravity in four-dimensional (4D) AFS may be dual to a theory living on the ``celestial sphere'' at infinity \cite{Pasterski:2016qvg}. This proposal is backed up by evidence that the Lorentz symmetry of scattering in 4D AFS is enhanced to Virasoro \cite{Kapec:2014opa}, as well as the existence of a stress tensor constructed from a particular subleading soft graviton mode in the bulk \cite{Kapec:2016jld}. In section \ref{sthms} will see how this follows from the subleading soft graviton theorem. In section \ref{campl} we formulate scattering in AFS in terms of a new observable: the celestial amplitude. We show how celestial amplitudes re-express the $\mathcal{S}$-matrix in a basis of boost eigenstates \cite{Pasterski:2016qvg,Pasterski:2017kqt}. (In contrast, conventional scattering amplitudes are computed in a plane wave basis.) Such a construction exists for scattering of both massive and massless particles which is illustrated with a calculation of the tree-level celestial amplitude of two massless and one massive scalars \cite{Lam:2017ofc}. In section \ref{csymm} we describe some recent developments centered around the theme of celestial symmetries. We show that both bulk translation symmetry, as well as the soft theorems imply the existence of celestial currents which constrain the celestial amplitudes. We will see for example that Poincar\'e symmetry can be used to completely fix celestial three-point functions and constrain four-point functions \cite{Stieberger:2018onx, Law:2019glh}, while subleading and subsubleading soft theorems can be used to completely fix the leading OPE coefficients in a collinear expansion of gluons and gravitons \cite{Pate:2019lpp}. We finally show that celestial theories contain an infinity of soft currents and compute their algebra in some examples \cite{Guevara:2021abz, Strominger:2021lvk}.

We have tried to give a self-contained overview of this rapidly growing field by choosing a particular path through the subject. Many fascinating recent developments have been left out. Explicit constructions of tree-level celestial amplitudes have appeared in \cite{Pasterski:2016qvg, Pasterski:2017ylz, Schreiber:2017jsr, Stieberger:2018edy, Kalyanapuram:2020aya, Brandhuber:2021nez, Jiang:2021xzy, Hu:2021lrx}. Loop corrections were addressed in \cite{Banerjee:2017jeg, Gonzalez:2020tpi, PhysRevD.102.126020} while properties of bulk scattering such as the double copy and connections to ambitwistor strings have been worked out in \cite{Adamo:2019ipt, Casali:2020vuy, Casali:2020uvr, Campiglia:2021srh}. Celestial symmetries in both gravity and gauge theories, as well as their constraints on celestial amplitudes have been discussed in \cite{Distler:2018rwu, Banerjee:2018gce, Stieberger:2018onx, Banerjee:2019aoy,Law:2019glh, Puhm:2019zbl, Fan:2020xjj, Fotopoulos:2019vac, Law:2020tsg, Ebert:2020nqf, Banerjee:2020vnt, Banerjee:2020kaa, Guevara:2021abz, Banerjee:2021cly, Pasterski:2021dqe, Pasterski:2021fjn, Strominger:2021lvk}. Analytic properties of celestial four-point functions in the complex boost-weight plane have been worked out in \cite{Arkani-Hamed:2020gyp, Chang:2021wvv} and conformal block expansions were computed in \cite{Nandan:2019jas, Law:2020xcf, Fan:2021isc, Atanasov:2021cje}. Infrared divergences and related aspects were discussed in \cite{Choi:2019sjs, Laddha:2020kvp, Hijano:2020szl, Himwich:2020rro, Hirai:2020kzx, Freidel:2021wpl, Gonzalez:2021dxw, Nguyen:2021qkt, Lippstreu:2021avq}. We hope these lectures provide a bridge between the earlier developments reviewed in \cite{Strominger:2017zoo} and more recent results.

\section{Soft theorems and asymptotic symmetries}
\label{sthms}
Soft theorems arise as conservation laws associated with asymptotic/large gauge symmetries \cite{as1, He:2014laa, He:2014cra, He:2015zea}. In this section we illustrate this connection by studying the (tree-level) subleading soft graviton theorem and the implied Virasoro symmetry of the $\mathcal{S}$-matrix. This section is a review of \cite{Kapec:2014opa} and \cite{Kapec:2016jld}.

\subsection{Soft theorems}

\begin{figure}
\begin{center}
\includegraphics[scale=0.22]{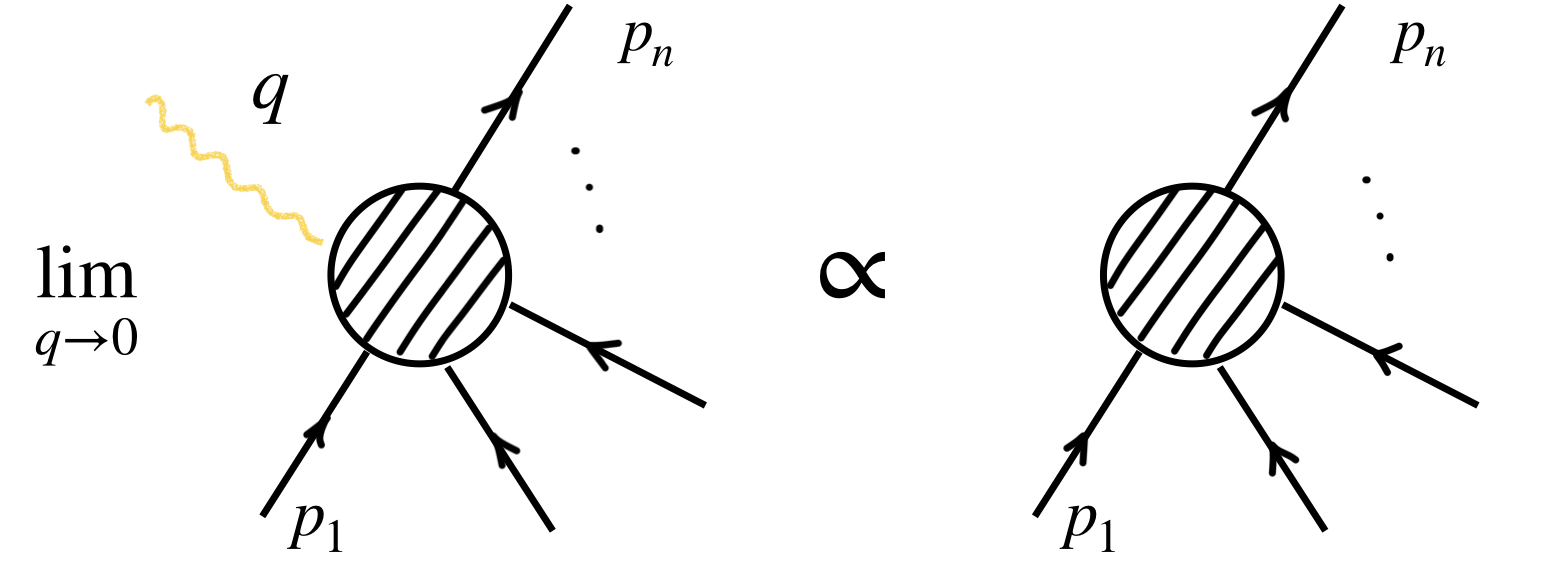}
\end{center}
\caption{The soft limit relates an amplitude with a low-energy massless particle to the same amplitude without the massless particle.}
\label{fig:soft-thm}
\end{figure}

We start by introducing a universal\footnote{Universal here  means independent of the nature of other particles involved in the scattering process.} relation obeyed by scattering amplitudes in any theory with massless particles. For simplicity, we focus on tree-level scattering. In gravity and gauge theory, the scattering of high-energy charged particles is accompanied by radiation. The radiation can be described as a collection of quanta (photons, gravitons, ...) of different energies. When the energy carried away by one such quantum is small, the scattering amplitude factorizes \cite{Low:1958sn, Weinberg:1965nx}
\begin{equation}
\label{soft-relation}
\lim_{\omega \rightarrow 0} \mathcal{A}_{n+1}^{\pm}(q) = \left[S_n^{(0)\pm} + S_n^{(1)\pm} + \mathcal{O}(\omega) \right]\mathcal{A}_n.
\end{equation}
Here $\mathcal{A}_{n+1}$ is a scattering amplitude of $n$ generic particles  of four-momenta $p_1,..., p_n$ and one massless particle of four-momentum $q = (\omega, \vec{q})$ and positive or negative helicity. $\mathcal{A}_n$ is the same scattering amplitude in the absence of the massless particle. This limit is illustrated in figure  \ref{fig:soft-thm} and will be referred to as the soft limit. 

$S^{(0)\pm}_n$ and $S^{(1)\pm}_n$ are the leading and subleading soft factors respectively, which take the form  \cite{Weinberg:1965nx, Low:1958sn,Cachazo:2014fwa} \vspace{-12pt}
\begin{equation}
\label{gr-sf}
\begin{split}
S_n^{(0)\pm} = \frac{\kappa}{2} \sum_{k=1}^n \frac{\left( p_k \cdot  \ve^\pm(q) \right)^2}{ p_k \cdot q }\ , \quad S_n^{(1)\pm} = - \frac{i\kappa}{2} \sum_{k=1}^n \frac{ \ve^{\pm} (q)\cdot p_k  }{ p_k \cdot q } q \cdot  \CJ_k \cdot \ve^{\pm}(q) \ , \quad \kappa = \sqrt{32 \pi G}
\end{split}
\end{equation}
in gravity\footnote{In gravity there is also a sub-subleading soft theorem \cite{Cachazo:2014fwa} with $S_n^{(2)\pm} = -\frac{\kappa}{4}\sum_{k = 1}^n \frac{(q\cdot J_k \cdot \varepsilon^{\pm})^2}{q \cdot p_k}$. } and
\begin{equation}
\label{gt-sf}
S^{(0)\pm}_n = \sum_{k=1}^n Q_k \frac{ p_k \cdot \ve^\pm (q)  }{ p_k \cdot q }\ , \quad S_n^{(1)\pm} = -i  \sum_{k=1}^n Q_k \frac{q \cdot \CJ_k \cdot \ve^{\pm}(q)}{ p_k \cdot q } 
\end{equation}
in quantum electrodynamics. $G$ and $Q_k$ are Newton's constant and the charges of the $n$ particles respectively. We expressed the polarization tensor $\ve_{\mu\nu}^{\pm} $ of  the graviton as \footnote{We pick a gauge in which the graviton is transverse and traceless, $q^{\mu} \varepsilon_{\mu\nu} = q^{\nu}\varepsilon_{\mu\nu} = \varepsilon^{\mu}_{\ \mu} = 0.$}
\begin{equation}
\varepsilon_{\mu\nu}^{\pm}(q) = \ve^\pm_\mu(q) \ve^\pm_\nu(q),
\end{equation}
where $\ve^\pm_{\mu}(q)$ is the polarization of a helicity-1 particle obeying 
\begin{equation}
\ve^\pm( q) \cdot q = 0 , \qquad  \ve^\pm( q)\cdot \ve^\pm ( q) = 0 , \qquad \ve^\pm ( q) \cdot \ve\,^\mp ( q) = 1. 
\end{equation}
$\CJ_k$ is the total angular momentum of particle $k$. For simplicity, we will work in units where
\be 
8\pi G = 1, \quad \kappa = \sqrt{32\pi G} = 2.
\ee

\begin{figure}
\begin{center}
\includegraphics[scale=0.23]{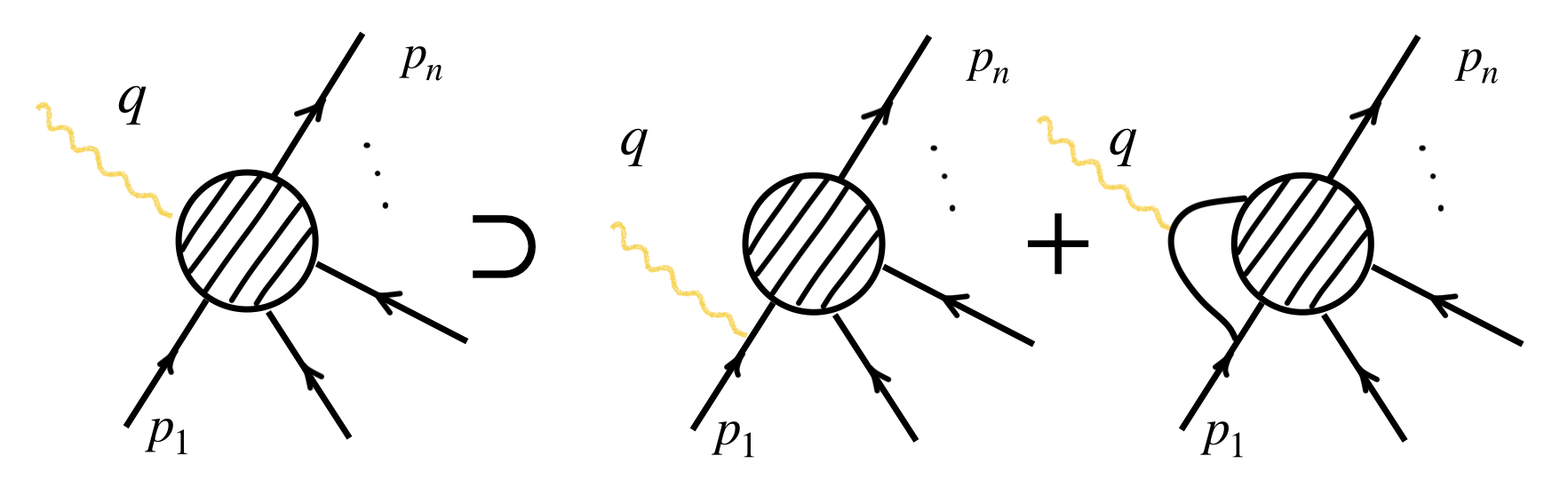}
\end{center}
\caption{In the soft limit, the amplitude will include contributions from Feynman diagrams where the soft particle attaches to external and internal lines. Diagrams where the soft particle attaches to an internal line are subleading in the soft limit.}
\label{fig:soft-proof}
\end{figure}

Notice that the soft theorem \eqref{soft-relation} captures the behavior of the scattering amplitude in an expansion around $\omega = 0$. The leading term in \eqref{soft-relation} has a simple pole at $\omega = 0$ which can be understood by considering the Feynman diagrams contributing to the scattering of $n+1$ particles,  as shown in figure \ref{fig:soft-proof}. In particular, as $\omega \rightarrow 0$, the leading order contribution comes from diagrams where the massless particle attaches to an external line. In this limit, an internal propagator goes on-shell and the amplitude develops a pole in $q$
\begin{equation}
\lim_{\omega \rightarrow 0}  \mathcal{A}_{n+1}(q) = \left[ \sum_{k = 1}^n -i\dfrac{V_k(\varepsilon, p_k)}{2p_k\cdot q} + \mathcal{O}(\omega^0) \right] \mathcal{A}_n,
\end{equation}
where $V_k(\varepsilon, p_k)$ is the leading term as $\omega \rightarrow 0$ in the (momentum-space) coupling at vertex $k$.
The remaining diagrams, where the massless particle attaches to an internal line remain finite as $\omega \rightarrow 0$. 

The analysis of subleading terms in the (tree-level) soft expansion was carried out explicitly in gauge theory \cite{Low:1958sn} and, more recently in gravity using on-shell amplitudes techniques \cite{Cachazo:2014fwa}. The ``brute-force'' computation is lengthy and subtle,\footnote{This is not only because of many sources of subleading corrections coming from both classes of diagrams in figure \ref{fig:soft-proof}, but also because at subleading order in the soft expansion, momenta of other external particles have to be deformed to obey overall momentum conservation.} yet a number of apparently miraculous cancellations yield the final result \eqref{soft-relation} universal in gravity \cite{He:2014bga,Elvang:2016qvq},\footnote{The subleading soft photon theorem may receive non-universal corrections from a short list of operators \cite{Elvang:2016qvq,Laddha:2017vfh}.} with the subleading soft graviton factor taking the simple form in \eqref{gr-sf}. 

It is often the case in physics that simple answers found as a result of  complicated calculations point towards an underlying symmetry of the theory. Indeed, as we will review in section \ref{Virasoro}, the subleading soft graviton theorem is nothing but a consequence of an infinite-dimensional enhancement of the Lorentz symmetry of the $\mathcal{S}$-matrix \cite{Kapec:2014opa}. Moreover, a certain mode of the soft graviton will be identitfied with the generator of this symmetry in section \ref{2dst} by recasting the subleading soft graviton theorem as the Virasoro-Ward identities of an insertion of the stress-tensor in a 2D conformal correlation function \cite{Kapec:2016jld}.

\subsection{Penrose diagram of Minkowski space}
\label{Penrose} 

Penrose diagrams are a convenient tool for studying physics at ``infinity'' as they preserve the causal structure of spacetime while mapping infinity to the boundary of a finite region. In this section we review how this works for Minkowski spacetime.

The Minkowski metric takes the form
\begin{equation}
\label{Mink}
ds^2 = -dt^2 + d\vec{x}^2 = -dt^2 + dr^2 + r^2 d\Omega_2^2,
\end{equation}
where
\begin{equation}
d\Omega_2^2 = d\theta^2 + (\sin \theta)^2 d\varphi^2 
\end{equation}
is the metric on the unit two-sphere. It will be convenient to introduce retarded and advanced coordinates $u, v$
\begin{equation}
u = t - r, \qquad v = t + r,
\end{equation}
and coordinates $(z, \bz)$ related to the angular coordinates $(\theta, \phi)$  by  a stereographic projection
\begin{equation}
z = \cot \frac{\theta}{2} e^{i\varphi}, \qquad \bz = \cot \frac{\theta}{2} e^{-i \varphi}.
\end{equation}
In retarded coordinates $(u, r, z, \bz)$ the metric \eqref{Mink} becomes
\begin{equation} 
\label{mink-scri+}
ds^2 = -du^2 - 2 du dr + 2r^2 \gamma_{z\bz} dz d\bz, \qquad \gamma_{z\bz} = \dfrac{2}{(1 + z\bz)^2}
\end{equation}
and similarly, in advanced coordinates $(v, r, z, \bz)$
\begin{equation}
\label{mink-scri-}
ds^2 = -dv^2 + 2 dv dr + 2r^2 \gamma_{z\bz} dz d\bz.
\end{equation}

\begin{figure}
\begin{center}
\includegraphics[scale=0.2]{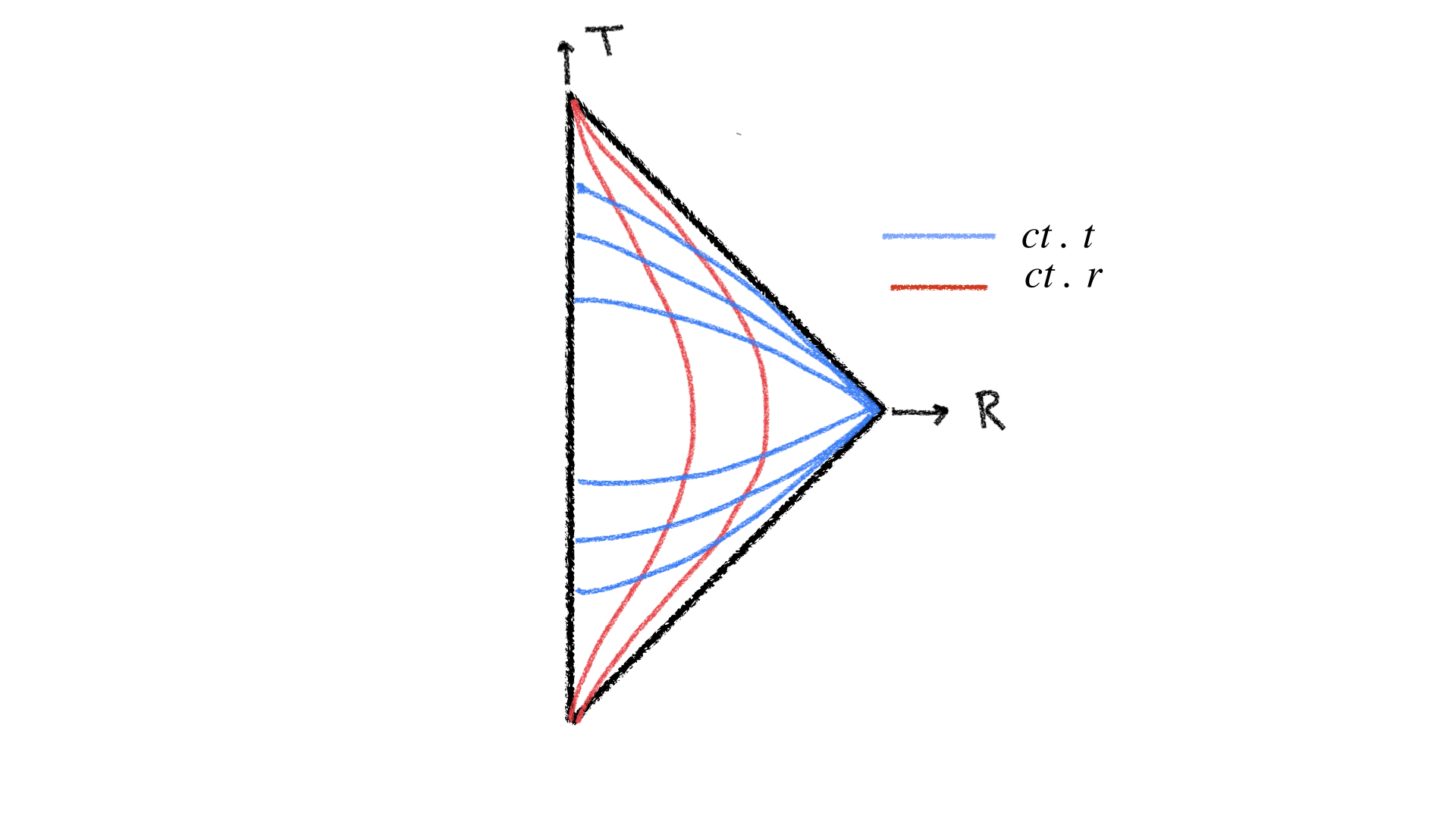}
\caption{Penrose diagram of Minkowski space.}
\label{fig:pd}
\end{center}
\end{figure}

The asymptotic structure of \eqref{Mink} can be understood by introducing coordinates $(T,R)$ related to $(t, r)$ by
\begin{equation}
\label{comp}
u = \tan U, \quad v = \tan V, \qquad T = U + V, \quad R = V - U,
\end{equation}
in which case \eqref{Mink} reduces to 
\begin{equation}
\begin{split}
ds^2 &= \Omega^2(T,R)\left(- dT^2 + dR^2 + 2 \sin^2 R \gamma_{z\bz} dz d\bz \right), \\
 \Omega^{-2}(T, R) &= 4 \cos^2\frac{1}{2}(T - R) \cos^2\frac{1}{2}(T + R).
\end{split}
\end{equation}
In the original coordinates, Minkowski space is covered by $r > 0$,  $-\infty < u < v < \infty$, therefore the ranges of the new coordinates are $-\frac{\pi}{2}< U < V < \frac{\pi}{2}$ and $0< R < \pi$. This is illustrated in figure \ref{fig:pd}. 

\begin{exercise}
\label{ct-lines}
\begin{enumerate}
\item[a)] Plot the lines of constant $r$ and $t$ in the $(R, T)$ plane.
\item[b)] Plot the lines of $\frac{r - r_0}{t}$ for different values of $r_0$ in the $(R, T)$ plane.
\end{enumerate}
\end{exercise}

\begin{figure}
\begin{center}
\includegraphics[scale=0.2]{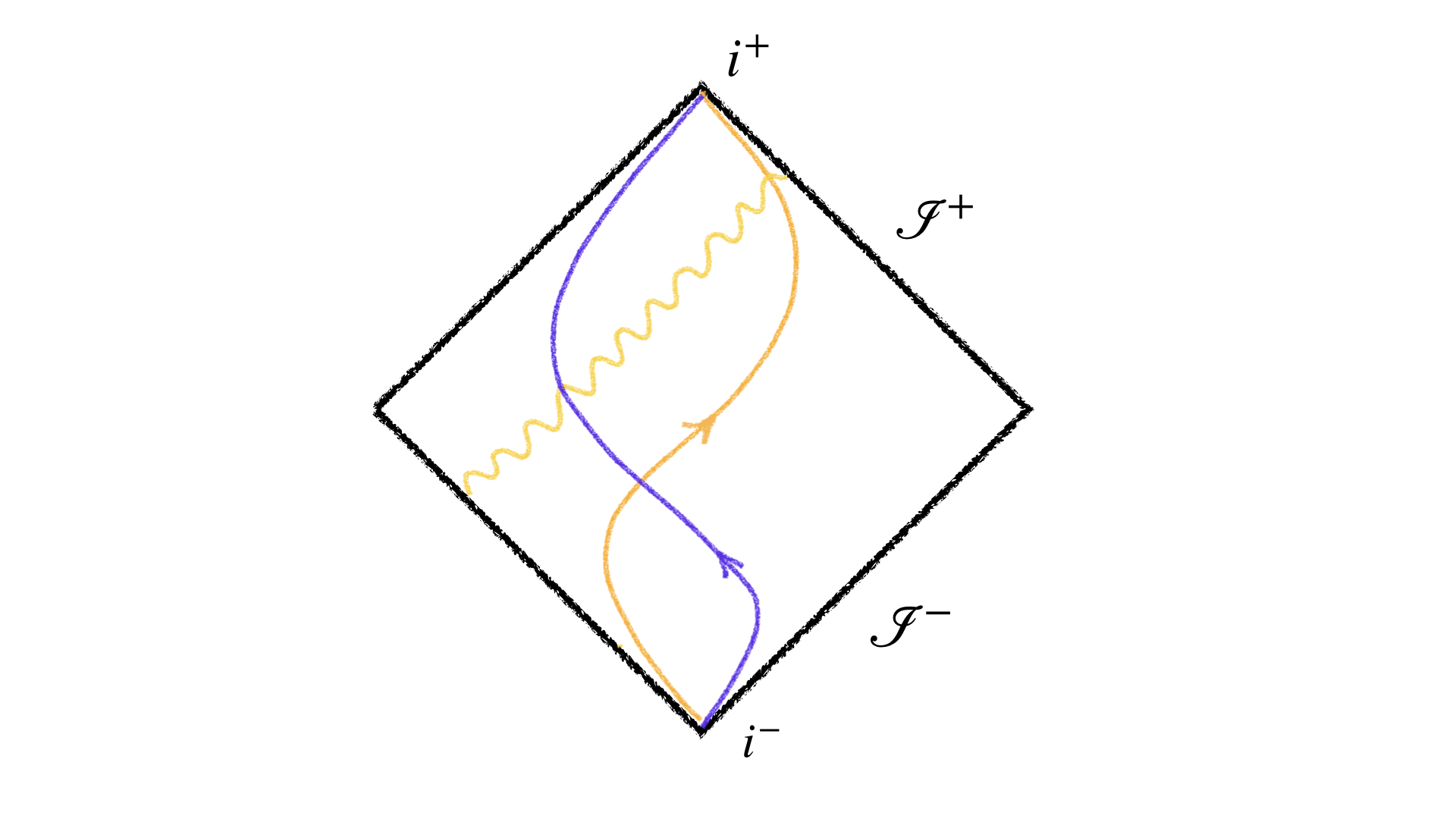}
\caption{Penrose diagram of Minkowski space where each pair of points represents a two-sphere. Massive particles come in from $i^-$ and go out at $i^+$, while massless particles enter and exit spacetime at $\mathcal{I}^{\pm}$.}
\label{fig:pd1}
\end{center}
\end{figure}

It will be convenient to unfold this diagram to represent antipodal points on the spheres. Future null infinity ($\mathcal{I}^+$) is defined by taking $r \rightarrow \infty$ for fixed $u$, while past null infinity ($\mathcal{I}^-$) is reached by taking $r\rightarrow \infty$ for fixed $v$. In a free theory, massless particles follow lines of unit slope and cross points on the spheres at $\mathcal{I}^{\mp}$ at retarded times $v, u$. This is illustrated in figure \ref{fig:pd1}. Massive particles never reach $\mathcal{I}^{\pm}$, but only past and future timelike infinities $i^{\mp}$ ($t \rightarrow \mp\infty$). 

\subsection{Asymptotically flat spacetimes}

Asymptotically flat spacetimes have the same causal structure as Minkowski space at infinity.
An asymptotically flat spacetime admits an expansion in powers of $r^{-1}$ around the Minkowski metric \eqref{mink-scri+} near $\mathcal{I}^+$\footnote{We are working in Bondi gauge defined by $\p_r {\rm det}\left( \frac{g_{AB}}{r^2}\right) = 0$ and $g_{rr} = g_{rA} = 0$ where $A, B$ run over the transverse indices $z, \bz.$ In these coordinates gravitational waves propagate radially outwards (equivalently, lines of constant $u, z, \bz$ are null) and the wavefronts are spherical.}
\be 
\label{mink-exp}
\begin{split}
ds^2 = &-du^2 - 2 du dr + 2r^2 \gamma_{z\bz} dz d\bz, \qquad \\
&+ \frac{2m_B}{r} du^2 + rC_{zz} dz^2 + r C_{\bz\bz} d\bz^2 + 2g_{uz} du dz + 2g_{u\bz} du d\bz + \cdots .
\end{split}
\ee
Solving the Einstein equations\footnote{We set $\kappa = 2$.} 
\be 
\label{EE}
R_{\mu\nu} - \frac{1}{2} g_{\mu\nu} R = T^M_{\mu\nu}
\ee
order by order in a large-$r$ expansion\footnote{The leading terms in the $uu$ and $uz$ components of the matter stress tensor are taken to be $\mathcal{O}(r^{-2})$.} one finds \cite{Barnich:2011mi}
\be 
g_{uz} = \frac{1}{2} D^z C_{zz} + \frac{1}{6r} C_{zz} D_z C^{zz} + \frac{2}{3r} N_z + \mathcal{O}(r^{-2}),
\ee
where $D_z$ is the covariant derivative associated with $\gamma_{z\bz}$. Here $m_B$ and $N_z$ are the Bondi mass aspect and angular momentum aspect respectively, while 
\be 
N_{zz} = \p_u C_{zz}
\ee
is the outgoing news tensor. They are all functions of $(u, z, \bz)$.

$m_B, C_{zz}, N_z$ are not all independent. They are related by constraint equations.\footnote{These are the components of \eqref{EE} along the tangent to $\mathcal{I}^+.$} The $uu$ constraint gives \cite{Barnich:2011mi}
\be 
\label{bmc}
\p_u m_B = \frac{1}{4} D_z^2 N^{zz} + \frac{1}{4} D_{\bz}^2 N^{\bz \bz} -\frac{1}{2} T_{uu}^{M (2)} - \frac{1}{4} N_{zz} N^{zz},
\ee
while the $uz$ constraint reduces to 
\be 
\label{amc}
\begin{split}
\p_u N_z = &\frac{1}{4}D_z\left(D_z^2 C^{zz} - D_{\bz}^2 C^{\bz\bz} \right) - T_{uz}^{M (2)} + \p_z m_B + \frac{1}{16} D_z \p_u\left(C_{zz} C^{zz} \right)\\
&-\frac{1}{4} \left(N^{zz} D_z C_{zz} + N_{zz} D_z C^{zz}\right) - \frac{1}{4}D_z\left(C^{zz} N_{zz} - N^{zz} C_{zz} \right).
\end{split}
\ee
We defined
\be 
T_{\mu\nu}^{M(2)} = \lim_{r\rightarrow \infty} r^2 T_{\mu\nu}^M.
\ee
\begin{exercise}
**Optional** Verify \eqref{bmc} and \eqref{amc}.
\end{exercise}
The square of $N_{zz}$ measures the energy carried by gravitational radiation. We learn from \eqref{bmc} that the Bondi mass (the integrated Bondi mass aspect over the sphere) is roughly speaking a measure of the net energy contained in spacetime excluding the parts carried off to infinity by null matter and gravitational waves: as energy is radiated away, the Bondi mass decreases. Similarly, the change in $N_z$ with retarded time (integrated over the sphere) measures the amount of angular momentum carried away to infinity by null matter and gravitational radiation. 

The analogous equations near $\mathcal{I}^{-}$ can be found in \cite{Kapec:2014opa}.

\subsection{Asymptotic symmetries}
The asymptotic symmetry group of \eqref{mink-exp} has beeen proposed to be the extended BMS$^+$ group in \cite{Barnich:2011mi}. This is generated by vector fields $\xi^+$ that preserve the asymptotic fall-off of \eqref{mink-exp} with $r$ namely
\be 
\label{sym}
\mathcal{L}_{\xi^+} g_{ur} = \mathcal{O}(r^{-2}), \quad \mathcal{L}_{\xi^+} g_{uz} = \mathcal{O}(1), \quad \mathcal{L}_{\xi^+} g_{zz} = \mathcal{O}(r), \quad \mathcal{L}_{\xi^+} g_{uu} = \mathcal{O}(r^{-1}).
\ee
Here $\mathcal{L}_{\xi^+}$ is the Lie derivative with respect to $\xi^+$. Solving \eqref{sym} order by order in a large-$r$ expansion, such vector fields are found to be of the form \cite{Barnich:2011mi, Kapec:2014opa}
\be 
\label{vf}
\begin{split}
\xi^+ = &\left(1 + \frac{u}{2r} \right) Y^{+z} \p_z  - \frac{u}{2r} D^{\bz} D_z Y^{+z} \p_{\bz} - \frac{1}{2}(u + r) D_z Y^{+z} \p_r + \frac{u}{2} D_z Y^{+z} \p_u + c.c.\\
&+ f^+\p_u - \frac{1}{r}\left(D^z f^+ \p_z + D^{\bz} f^+ \p_{\bz} \right) + D^zD_z f^+ \p_r + \cdots,
\end{split}
\ee
where $f^+(z, \bz)$ is an arbitrary function on $\mathcal{S}^2$ and $Y^+(z, \bz)$ is a conformal Killing vector on $\mathcal{S}^2$
\be 
\label{cke}
\p_{\bz} Y^{+ z} = 0.
\ee
\begin{exercise}
Use \eqref{sym} and \eqref{mink-exp} to derive \eqref{vf}.
\end{exercise}
One easy way to see that \eqref{cke} ought to hold is to notice that under Lorentz transformations,
\be 
\mathcal{L}_{Y^+}g_{\bz\bz} = 2r^2 \gamma_{z\bz} \p_{\bz} Y^{+ z} + \mathcal{O}(r).
\ee
Imposing that \eqref{sym} are obeyed immediately leads to \eqref{cke}.

Globally, \eqref{cke} admits six solutions 
\be 
\label{gkvf}
\begin{split}
Y_{12}^{+z} &= -iz, \quad Y^{+z}_{13}=  \frac{1}{2}(1 + z^2), \quad Y^{+z}_{23} = \frac{i}{2}(1 - z^2), \\
Y_{03}^{+z} &= -z, \quad Y^{+z}_{01}=  \frac{1}{2}(1 - z^2), \quad Y^{+z}_{02} = \frac{i}{2}(1 + z^2)
\end{split}
\ee
corresponding to the three Lorentz rotations and three boosts (see appendix \ref{mwcp}). Locally, there are infinitely many solutions $Y^{+z} \propto z^n.$

For the remainder of this section, we restrict to the subgroup of asymptotic symmetries generated by \eqref{vf} with $f = 0$ which are known as superrotations \cite{Barnich:2011ct}.\footnote{Conversely, supertranslations are the subset of symmetries \eqref{vf} with $Y^+ = 0$  and $f \neq 0$.} In this case, the vector fields \eqref{vf} that map $\mathcal{I}^+$ to itself are
\be 
\label{vfscri}
\xi^+\Big|_{\mathcal{I}^+} =  Y^{+z} \p_z  + \frac{u}{2} D_z Y^{+z} \p_u + c.c. .
\ee
The infinitesimal BMS$^+$ transformations \eqref{vfscri} act on the metric components as follows
\be
\label{symm-action} 
\begin{split}
\delta_{Y^+} C_{zz} &= \frac{1}{2}\left(D_z Y^{+z} + D_{\bz} Y^{+\bz} \right)\left(u\p_u - 1 \right)C_{zz} + \mathcal{L}_{Y^+} C_{zz} - u D_z^3 Y^{+z}, \\
\delta_{Y^+} N_{zz} &= \p_u \delta_{Y^+} C_{zz} = \frac{u}{2}\left(D_z Y^{+z} + D_{\bz} Y^{+\bz} \right)\p_u N_{zz} + \mathcal{L}_{Y^+} N_{zz} - D_z^3 Y^{+z}.
\end{split}
\ee
 Upon quantization, \eqref{symm-action} imply the existence of ``charges''\footnote{We haven't shown these are conserved yet; conservation will be implied by the subleading soft graviton theorem.}
under which an outgoing Fock state\footnote{The radiative data consists of the modes of $N_{zz}$ hence the action of $Q^+$ on the radiative part of $|{\rm out}\rangle$ is related to the commutator $[Q^+, N_{zz}]$  \cite{He:2014laa}.} transforms as 
\be 
Q^+(Y^+)|{\rm out} \rangle = i \delta_{Y^+} |{\rm out} \rangle,
\ee
where 
\be 
\label{ch}
Q^+(Y^+) = Q^+_H + Q^+_S
\ee
and \cite{Barnich:2011ct}
\be 
\label{charges}
\begin{split}
Q_H^+ &= \frac{1}{4}\int_{\mathcal{I}^+} du d^2z \gamma_{z\bz} \left( u D_z Y^{+z} N_{zz} N^{zz}  - Y^{+z} D_z(C_{zz} N^{zz}) - 2 Y^{+z} C_{zz} D_z N^{zz} +{\rm matter} \right)\\
&\hspace{370pt} + c.c., \\
Q_S^+ &= -\frac{1}{2}\int_{\mathcal{I}^+} du d^2z D_z^3 Y^{+z} u N^{z}_{\ \bz} + c.c..
\end{split}
\ee
We show in appendix \ref{commutation-rel} that \eqref{charges} reproduce the symmetry action \eqref{symm-action}.
Using the canonical commutation relations \cite{Ashtekar:1987tt} 
\be 
[N_{\bz\bz}(u, z, \bz), C_{ww}(u', w, \bw)] = 2i \gamma_{z\bz} \delta^{(2)}(z - w) \delta(u - u'),
\ee
it then follows that for transformations parameterized by $Y^+ = (Y^{+z}, 0)$ \cite{Kapec:2014opa}
\be 
\label{hcaction}
Q_H^+|{\rm out}\rangle = i \sum_{k \in {\rm out}}\left(\mathcal{L}_{Y^{+z_k}} - \frac{\omega_k}{2} D_{z_k}Y^{+z_k}\p_{\omega_k} \right)|{\rm out}\rangle.
\ee
Similar formulas hold near $\mathcal{I}^-.$

\subsection{Recovering the Virasoro symmetry from the soft theorem}
\label{Virasoro}

An independent action of BMS$^+$ and BMS$^-$ on $\mathcal{I}^+$ and $\mathcal{I}^-$ leads to an ambiguity in defining scattering in AFS. In particular, upon specifying data at $\mathcal{I}^+,$ the $S$-matrix provides a map between in and out states \textit{up to} a BMS transformation. 
 
A solution to this problem was proposed in \cite{as1} where it was shown that the gravitational scattering problem in AFS becomes well defined upon imposing the antipodal matching conditions
\be 
f(z, \bz) \Big|_{\mathcal{I}^+_-} = f(z, \bz) \Big|_{\mathcal{I}^-_+}, \quad Y^+(z, \bz) \Big|_{\mathcal{I}^+_-} = Y^-(z, \bz) \Big|_{\mathcal{I}^-_+}.
\ee
Here, points on the sphere at $\mathcal{I}^+_-$ are antipodally related to points at $\mathcal{I}^-_+$, $(z, \bz)\Big|_{\mathcal{I}^+_-} = (-\frac{1}{\bz}, -\frac{1}{z})\Big|_{\mathcal{I}^-_+}$.
Moreover, upon imposing the boundary condition\footnote{Using the constraint \eqref{amc} the superrotation charges can be put into the simpler form $Q^+(Y^+) = \int d^2 z \left(Y_{\bz}^+ N_z + Y_z^+ N_{\bz} \right)$ \cite{Strominger:2017zoo}.} 
\be 
N_z(z, \bz)\Big|_{\mathcal{I}^+_-} = N_z(z, \bz)\Big|_{\mathcal{I}^-_+}, 
\ee
the charges at $\mathcal{I}^-_+$ and $\mathcal{I}^+_-$ obey
\be 
Q^+ = Q^-.
\ee
The conservation of $Q$ implies the $\mathcal{S}$-matrix obeys the constraint
\be 
\label{cl}
\langle {\rm out}| Q^+ \mathcal{S} - \mathcal{S} Q^- | {\rm in}\rangle = 0.
\ee
Using the split \eqref{ch} into soft and hard parts, \eqref{cl} is equivalent to
\be 
\label{st}
\langle {\rm out}| Q^+_S \mathcal{S} - \mathcal{S} Q^-_S | {\rm in}\rangle = -\langle {\rm out}| Q^+_H \mathcal{S} - \mathcal{S} Q^-_H | {\rm in}\rangle.
\ee
$Q_S$ is linear in the news and it can be shown to correspond to a mode of the graviton. In particular, one finds \cite{Kapec:2014opa}
\be 
N_{\bz \bz}^{(1)} = \int_{-\infty}^{\infty} du u N_{\bz\bz} = \frac{i}{4\pi} \hat{\epsilon}_{\bz \bz} \lim_{\omega \rightarrow 0} (1 + \omega \p_{\omega}) \left[a_-^{\rm out}(\omega \hat{x}) - a_+^{{\rm out}\dagger}(\omega \hat{x}) \right]
\ee
and hence $Q_S$ picks out a particular subleading\footnote{$1 + \omega \p_{\omega}$ projects out the leading soft pole.} soft graviton mode. 

For simplicity we now restrict to the case when all asymptotic particles but the soft insertion are scalars. Using the subleading soft relation \eqref{gr-sf}, the LHS of \eqref{st} reduces to \cite{Kapec:2014opa}
\be 
\label{ST}
\begin{split}
\langle {\rm out}| Q^+_S \mathcal{S} - \mathcal{S} Q^-_S | {\rm in}\rangle &= -\frac{i}{4\pi} \int d^2z D_z^3 Y^z \hat{\epsilon}_{\bz\bz} S^{(1)-} \langle {\rm out}| \mathcal{S}|{\rm in} \rangle\\
&=- i \sum_{k \in {\rm in, out}} \left(Y^{z_k} \p_{z_k} - \frac{\omega_k}{2} D_{z_k}Y^{z_k} \p_{\omega_k} \right) \langle {\rm out}|\mathcal{S}|{\rm in}\rangle.
\end{split}
\ee
In the second line, we integrated by parts and used the parameterizations of momenta 
\be 
\label{q}
q(\omega, z, \bz) =\omega \hat{q}(z, \bz) = \frac{\omega}{1 + z\bz} \left(1 + z\bz, z + \bz, -i(z - \bz), 1 - z\bz \right)
\ee 
for which the subleading soft factor for a negative helicity graviton becomes
\be 
S^{(1)-} = \sum_k \left(\frac{(z - z_k)(1 + z\bz_k)}{(\bz_k - \bz)(1 + z_k\bz_k)} \omega_k \p_{\omega_k} + \frac{(z - z_k)^2}{\bz_k - \bz}\p_{z_k} \right).
\ee
The RHS of \eqref{ST} coincides with the action \eqref{hcaction} of the hard charge $Q_H$ on scalar asymptotic states. We conclude the tree-level subleading soft theorem implies the conservation of the charges \eqref{ch}, hence the enhancement of Lorentz symmetry to Virasoro. 

\subsection{A 2D stress tensor for 4D gravity}
\label{2dst}

\eqref{ST} also points to another remarkable feature of gravity in AFS: the existence of a subleading soft graviton mode whose insertions in the quantum gravity $\mathcal{S}$-matrix behaves like the stress tensor of a 2D CFT. To see this, we simply set 
\be 
Y^{z_k} = \frac{1}{z - z_k}
\ee
in \eqref{ST}. Upon defining\footnote{Note that this operator is directly related to the soft charges $Q_S^{\pm}$ evaluated at $Y^w = \frac{1}{z - w}$.}
\be 
T_{zz} \equiv  i\int d^2 w \frac{1}{z - w} D_w^2 D^{\bw} N_{\bw \bw}^{(1)},
\ee
\eqref{ST} reduces to 
\be 
\label{stWi}
\langle T_{zz} \mathcal{O}_1... \mathcal{O}_n\rangle = \sum_{k = 1}^n \left[\frac{\hat{h}_k}{(z - z_k)^2}  + \frac{\Gamma_{\ z_k z_k}^{z_k}}{z - z_k}\hat{h}_k + \frac{1}{z - z_k}\p_{z_k} \right]\langle \mathcal{O}_1... \mathcal{O}_n\rangle.
\ee
This is the Ward identity of a stress tensor in a conformal field theory on a curved background \cite{EGUCHI1987308}. Note however that the weight\footnote{For external states of spin $s_k$, the weights generalize to $\hat{h}_k = \frac{s_k -\omega_k \p_{\omega_k}}{2}$ and \eqref{stWi} gets corrected by a spin connection term, see \cite{Kapec:2016jld} for the general formula.} $\hat{h}_k = -\frac{1}{2}\omega_k \p_{\omega_k}$ is a differential operator which acts non-diagonally on $S$-matrix elements in a basis of momentum eigenstates. In the next section we will introduce a new basis of asymptotic states which diagonalize the action of $\hat{h}_k$. The scattering problem in AFS will be then reformulated in terms of an observable living on the celestial sphere: the \textit{celestial amplitude}. 

\section{Celestial amplitudes}
\label{campl}

In this section we introduce a basis for scattering in AFS which diagonalizes asymptotic boosts as opposed to momentum generators. We show how to formulate scattering in this basis, with celestial amplitudes defining observables living on the sphere at infinity. We illustrate this construction by computing the celestial amplitude of two massless and one massive scalars. This section is based on \cite{Pasterski:2016qvg} and \cite{Pasterski:2017kqt}.

\subsection{Conformal primary wavefunctions}

Scalar conformal primary wavefunctions are solutions to the wave equation
\be 
\label{we0}
\left(\nabla^2 - m^2 \right) \Psi = 0,
\ee 
which are ``highest weight'' with respect to the Lorentz SO$(1,3)$. We start by identifying the associated highest weight states, then impose they are solutions to \eqref{we0}. 
A representation of the Lorentz generators is
\be 
\label{Lorentz-full}
J^{\mu\nu} = L^{\mu\nu} + S^{\mu\nu},
\ee
where
\be 
\label{Lorentz}
L^{\mu \nu} = -\left(x^{\mu} \p^{\nu} - x^{\nu} \p^{\mu} \right)
\ee
is the orbital angular momentum generator and $S^{\mu\nu}$ is the spin generator. For simplicity, we focus on scalars in which case $S^{\mu\nu} = 0.$

\eqref{Lorentz} consist of rotations
\begin{equation}
\label{rot}
J_1 = -(x^2\partial_{x^3} - x^3\partial_{x^2}),~~~ J_2 = x^1\partial_{x^3} - x^3\partial_{x^1}, ~~~ J_3 = -(x^1\partial_{x^2} - x^2\partial_{x^1})
\end{equation}
and boosts
\begin{equation}
\label{boosts}
\begin{split}
K_1 = -( x^0\partial_{x^1} + x^1\partial_{x^0}), ~~~ K_2 = -(x^0\partial_{x^2} + x^2\partial_{x^0}), ~~~ K_3 = -(x^0\partial_{x^3} + x^3\partial_{x^0}).
\end{split}
\end{equation}
These generators obey the standard Lorentz algebra
\be 
\label{clg}
\begin{split}
[J_i, J_j] &= \epsilon_{ijk} J_{k}, \\
[K_i, K_j] &= -\epsilon_{ijk} J_k, \\
[J_i, K_j] &= \epsilon_{ijk} K_k,
\end{split}
\ee
while the linear combinations \eqref{genn} of \eqref{rot}, \eqref{boosts} in appendix \ref{mwcp} obey the SL$(2, \mathbb{C})$ commutation relations
\be 
\label{clgp}
[L_m, L_n] = (m - n) L_{m + n}, \quad [\bar{L}_m, \bar{L}_n] = (m - n) \bar{L}_{m + n}.
\ee

We now notice that 
\be 
\label{cps0}
\Psi_{\Delta} \propto \frac{1}{(x^0 + x^3)^{\Delta}} 
\ee
obeys 
\be 
\label{ev0}
(L_0 + \bar{L}_0)\Psi_{\Delta} = \Delta \Psi_{\Delta}, \quad (L_0 - \bar{L}_0) \Psi_{\Delta} = 0
\ee
as well as
\be 
\label{hw0}
L_1 \Psi_{\Delta} = \bar{L}_1\Psi_{\Delta} = 0.
\ee
In other words, \eqref{cps0} diagonalizes boosts in the $x^3$ direction and obeys the highest weight condition \eqref{hw0}. 

One could repeat this analysis starting with a set of rotated bulk Lorentz generators, 
\be 
J_i' = R_{ij} J_j, \quad K_i' = R_{ij} K_j,
\ee
where 
\begin{equation*} 
R = \left( \begin{matrix}
\cos \hat{\varphi} \cos \hat{\theta} && \sin \hat{\varphi} \cos \hat{\theta} &&-\sin \hat{\theta} \\
-\sin\hat{\varphi} && \cos \hat{\varphi} && 0  \\
\cos \hat{\varphi} \sin \hat{\theta} && \sin \hat{\varphi} \sin \hat{\theta} && \cos \hat{\theta}
\end{matrix} \right)  \equiv \left( \begin{matrix}
\hat{n}_1 \\
\hat{n}_2  \\
\hat{n}_3
\end{matrix} \right).
\end{equation*}\break
Multiplication by an arbitrary function $f$ of the Lorentz invariant $x^2$ will preserve both the eigenvalue and highest weight conditions \eqref{ev0}, \eqref{hw0}, hence in general, a highest weight solution diagonal with respect to $K_3'$ will be\footnote{$(\hat{\theta}, \hat{\varphi})$ are related to $z, \bz$ via the stereographic projection.}
\be 
\label{hwa}
\Psi_{\Delta}(\hat{q}; x)  = \frac{f(x^2)}{(\hat{q} \cdot x)^{\Delta}}, \quad \hat{q} = \left(1, \hat{n}_3 \right) =  \hat{q}(z, \bz),
\ee
with $\hat{q}(z, \bz)$ in \eqref{q}.

Finally, we require that \eqref{hwa} obeys the wave equation. Plugging \eqref{hwa} into \eqref{we0}, we find the following differential equation for $f$ \cite{Pasterski:2017kqt}
\be 
\label{Be}
4x^2 f''(x^2) - 4(\Delta - 2) f'(x^2) - m^2 f(x^2) = 0.
\ee
The solutions to  \eqref{Be} are linear combination of Bessel functions (of first kind)
\be 
\label{sol0}
f(x^2)= \left(\sqrt{-x^2}\right)^{\Delta - 1} \left[c_1 I_{\Delta - 1}(m\sqrt{x^2}) + c_2 I_{-\Delta + 1}(m\sqrt{x^2})\right],
\ee
where $c_1, c_2$ are ($\Delta$-dependent) constants.
Imposing that \eqref{sol0} decays to $0$ as $x^2 \rightarrow \infty$ picks out the linear combination proportional to the Bessel function of second kind
\be
f(x^2) \propto \left(\sqrt{-x^2}\right)^{\Delta - 1} K_{\Delta - 1}(m \sqrt{x^2}). 
\ee
We conclude that up to normalization, the massive conformal primary wavefunctions take the form\footnote{An $i\epsilon$ prescription distinguishes between in and out states \cite{Pasterski:2017kqt}.}
\be 
\label{mcpw}
\Psi_{\Delta}(\hat{q};x) \propto \frac{ \left(\sqrt{-x^2}\right)^{\Delta - 1} }{(\hat{q}\cdot x)^{\Delta}}K_{\Delta - 1}(m \sqrt{x^2}). 
\ee

Under Lorentz transformations, both $\hat{q}$ and $x$ transform linearly
\be 
\begin{split}
\hat{q}^{\mu}(z', \bz') &= \left|\frac{\p \vec{z}'}{\p \vec{z}} \right|^{1/2} \Lambda^{\mu}_{\  \nu} \hat{q}^{\nu}(z, \bz), \\
x^{\mu'} &= \Lambda^{\mu}_{\  \nu} x^{\nu}
\end{split}
\ee
hence \eqref{mcpw} obeys
\be 
\label{sl2c}
\Psi_{\Delta}(\Lambda^{\mu}_{\ \nu} x; \vec{z}'(\vec{z})) = \Big|\frac{\p \vec{z}'}{\p \vec{z}} \Big|^{-\Delta/2} \Psi_{\Delta}(x; \vec{z}).
\ee
\begin{exercise}
Show that the conformal primary wavefunctions  obey \eqref{sl2c}.
\end{exercise}
In the next section we give an alternate derivation of \eqref{mcpw} which will lead to a representation of \eqref{mcpw} as Fourier transforms of $AdS_3$ bulk-to-boundary propagators.

\subsection{Milne slicing}

\begin{figure}
\begin{center}
\includegraphics[scale=0.2]{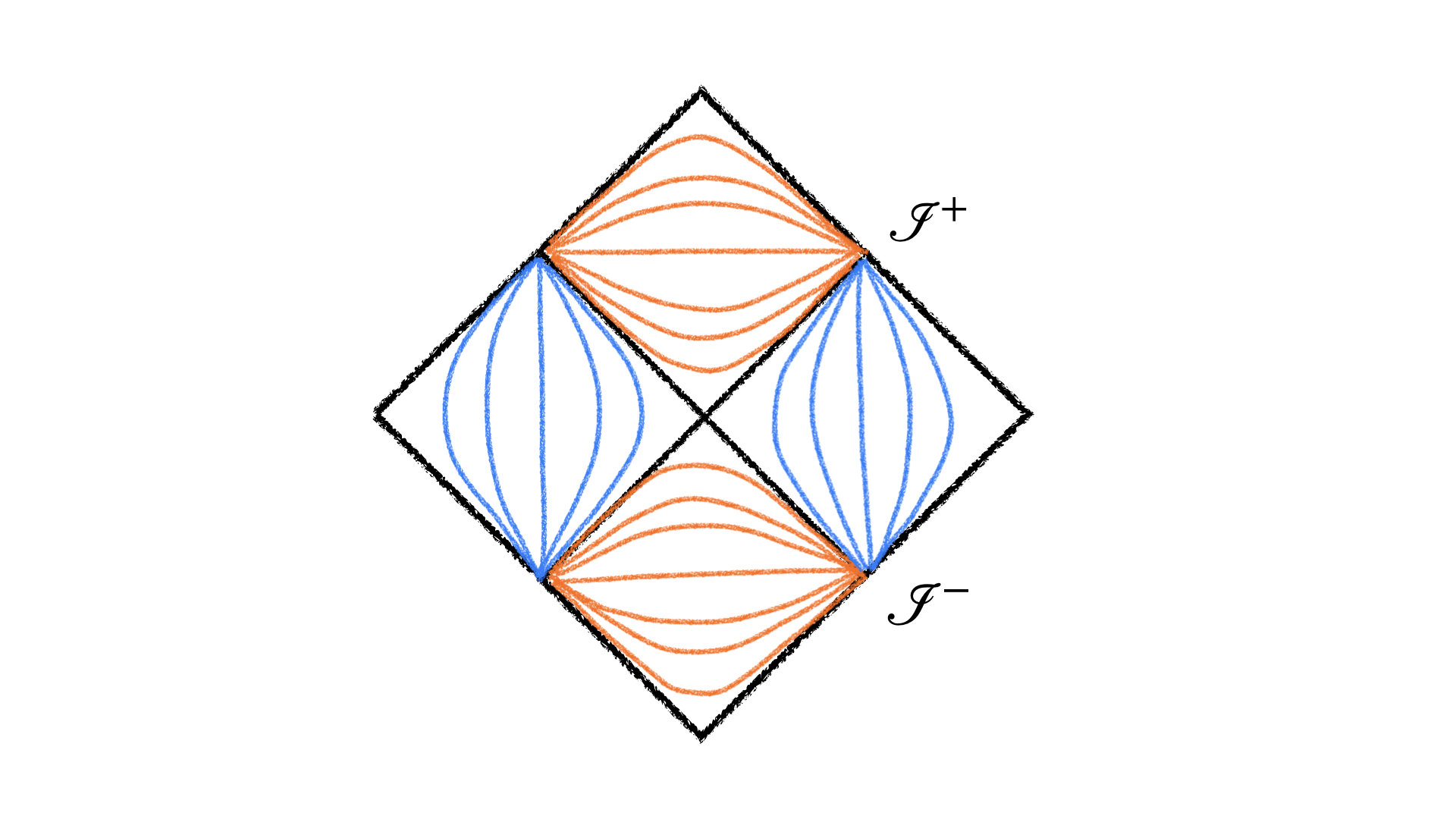}
\end{center}
\caption{Minkowski space split into four regions: the past and future lightcones are covered by $\mathbb{H}_3$ slices while the causally disconnected Rindler regions are covered by $dS_3$ slices.}
\label{fig:Milne-Rindler}
\end{figure}

The conformal compactification of Minkowski space in section \ref{Penrose} obscures one aspect of scattering in AFS: all massive particles enter (exit) spacetime at a point, $i^- (i^+)$, so how are we supposed to distingush between different asymptotics? The key is to resolve past and future timelike infinities by introducing the new coordinates \cite{deBoer:2003vf, Campiglia:2015qka}
\be 
\begin{split}
t^2 - r^2 &= \tau^2,\\
\rho \tau &= r.
\end{split}
\ee
In $(\tau, \rho, z, \bz)$ coordinates, \eqref{Mink} becomes
\be 
\label{H3}
ds^2 = -d\tau^2 + \tau^2 \left(\frac{d\rho^2}{1 + \rho^2} + 2 \rho^2 \gamma_{z\bz} dz d\bz \right) = -d\tau^2 + \tau^2 ds_{\mathbb{H}_3}.
\ee
We learn that slices of constant $\tau$ correspond to hyperboloids of radius $\tau$, while  $\rho  = \dfrac{r}{t}\left(1 - \dfrac{r^2}{t^2}\right)^{-1/2}$ is constant whenever $r/t$ is constant. Note also that since $t  = \tau \sqrt{1 + \rho^2}$, the limit $\tau \rightarrow \infty$ corresponds to $t\rightarrow \infty$ for fixed $(\rho,z, \bz)$. We illustrate the foliations of the past and future light-cones (also known as Milne wedges) with $\mathbb{H}_3$ slices in figure \ref{fig:Milne-Rindler}. 

Parameterizing the trajectory of a massive particle of constant momentum $\vec{p}$ and energy $E$ by 
\be 
\vec{r} = \vec{r}_0 + \frac{t}{E} \vec{p},
\ee
we find that as $t \rightarrow \infty$,
\be 
\rho \rightarrow \frac{|\vec{p}|}{m}, \quad \frac{\vec{r}}{r}\rightarrow \frac{\vec{p}}{p}.
\ee
Hence at late times, the massive particle asymptotes to fixed $(\rho, z, \bz)$, or equivalently a point on the unit hyperboloid at $i^+$.  Similarily, the Rindler wedges can be foliated with $dS_3$ slices. This slicing is easily obtained by letting
\be 
\tau = i \widetilde{\tau}, \quad \rho = -i \widetilde{\rho}
\ee
in \eqref{H3} in which case
\be 
\label{dSm}
ds^2 = d\widetilde{\tau}^2 + \widetilde{\tau}^2\left(\frac{d\widetilde{\rho}^2}{1 - \widetilde{\rho}^2} + 2\widetilde{\rho}^2\gamma_{z\bz} dz d\bz  \right).
\ee
This $dS_3$ slicing won't be discussed further herein, but see \cite{Liu:2021tif} for an analysis of associated conformal primary solutions. 

The proper, orthocronous Lorentz group acts as the group of isometries on the $\mathbb{H}_3$ slices for $t > 0$. To find solutions to \eqref{we0} that preserve slices of constant $\tau$ it is convenient to express $\nabla^2$ with respect to the coordinates  in \eqref{H3} .
\begin{exercise}
Using 
\be 
\nabla^2 \Psi = \frac{1}{\sqrt{g}} \p_{\mu} \left(g^{\mu\nu} \sqrt{g} \p_{\nu} \Psi \right)
\ee
show that in the coordinates \eqref{H3}, \eqref{we0} becomes
\be 
\label{weh3}
\left[\frac{1}{\rho \tau^2}\left(\left(3 \rho ^2+2\right) \p_{\rho} +\rho  \left(\rho ^2+1\right) \p_{\rho}^2 -\rho  \tau  \left(3 \p_{\tau}+\tau \p_{\tau}^2 \right)\right) + \frac{\Box_{S^2}}{\rho^2 \tau^2}\right] \Psi = m^2 \Psi.
\ee
\end{exercise}
Setting 
\be 
\rho = \sinh \eta,
\ee
\eqref{weh3} becomes
\be 
\label{we}
\left[\frac{1}{\tau^2}\left(\p_{\eta}^2 + 2\coth \eta \p_{\eta} \right) - 3\frac{\p_{\tau}}{\tau} - \p_{\tau}^2 + \frac{\Box_{S^2}}{\sinh^2 \eta\tau^2} \right] \Psi = m^2 \Psi.
\ee
%{\color{red} I think this is correct since it matches section 4 in Lowe upon $\tau \rightarrow i\tau$ and $\sinh \eta \leftrightarrow \cosh \eta$.}
This equation can be solved by separation of variables 
\be 
\Psi = \phi_{p, l}(\eta) \varphi_{p}(\tau) Y_{lm}(z, \bz),
\ee
where 
\be 
\label{wes}
\begin{split}
\left(\p_{\eta}^2 + 2\coth \eta \p_{\eta} + \frac{-l(l + 1)}{\sinh^2\eta} - p^2\right) \phi_{p,l}(\eta) = 0,\\
\left(-3\frac{\p_{\tau}}{\tau} - \p_{\tau}^2 + \frac{p^2}{\tau^2} - m^2\right) \varphi_{p}(\tau) = 0
\end{split}
\ee
and 
\be 
\Box_{S^2} Y_{lm} = -l(l+1) Y_{lm}.
\ee
Note that $p$ is a free parameter which cancels in \eqref{we}. We recognize the first equation in \eqref{wes} as the massive wave equation in AdS$_3$ while the second equation has two linearly independent solutions which can be taken to be
\be 
\varphi_{p}(\tau) = \frac{I_{\sqrt{1 + p^2}}(m\tau)}{\tau}, \quad \frac{K_{\sqrt{1 + p^2}}(m\tau)}{\tau}.
\ee
Choosing the second solution as it decays at $\tau \rightarrow \infty$, we recover the $\tau$ dependence in \eqref{mcpw} upon identifying 
\be 
\label{pd}
\Delta - 1 = \sqrt{1 + p^2}.
\ee
Using \eqref{pd} it is a standard exercise in AdS$_3$ to show that the first equation in \eqref{wes} can be written in terms of the SL$(2, \mathbb{C})$ generators \eqref{genn} (upon an appropriate coordinate transformation)
\be 
\label{ccas}
\left(4L_0^2 - 2L_{-1} L_1 - 2 L_{1} L_{-1} \right) \phi_{p, l} = \left(4\bar{L}_0^2 - 2\bar{L}_{-1} \bar{L}_1 - 2 \bar{L}_{1} \bar{L}_{-1} \right) \phi_{p, l}  = \Delta(\Delta - 2) \phi_{p, l}.
\ee
Using the SL$(2, \mathbb{C})$ commutation relations \eqref{clgp} and imposing that $L_{1} \phi_{p, l} = \bar{L}_1 \phi_{p,l} = 0$, \eqref{ccas} reduces to \eqref{ev0}, \eqref{hw0} and we recover the solutions \eqref{mcpw}. 

Note that \eqref{wes} admit more general solutions which don't obey the highest weight condition \eqref{hw0}. These can be used to construct the unitary principal series representations of SL$(2, \mathbb{C})$. This complementary calculation is detailed for the $dS_3$ slicing of Minkowski space \eqref{dSm} in \cite{Liu:2021tif}. A discussion of conformal primary solutions of \eqref{we0} in $(2,2)$ signature can be found in \cite{Atanasov:2021oyu}.

\subsection{Integral representation}

The conformal primary wavefunctions \eqref{mcpw} admit the Fourier representation
\be 
\label{intm}
\Psi_{\Delta}(x; \vec{z}) = \int_{\mathbb{H}_3} d^3\hat{p} G_{\Delta}(\hat{p}; \vec{z}) e^{im \hat{p}\cdot X},
\ee
where the momenta
\be
\label{massive-mom}
\hat{p}(y, w, \bw) =  \frac{1}{2y}\left(1 + y^2 + w\bw, w + \bw, -i(w - \bw), 1 - y^2 - w\bw \right)
\ee
are in one-to-one correspondence with points on the unit hyperboloid at $i^+$ and
\be 
\label{bbp}
G_{\Delta}(y, w, \bw; z, \bz) = \left(\frac{y}{y^2 + |z - w|^2} \right)^{\Delta}
\ee
is the bulk-to-boundary propagator in $AdS_3$ \cite{Witten:1998qj}. As they are weighted integrals of plane waves, they automatically solve the wave equation. That they transform as \eqref{sl2c} under SL$(2, \mathbb{C})$
follows from the transformation property 
\be 
G_{\Delta}(\hat{p}'; \hat{q}') = \left|\frac{\p \vec{z}'}{\p \vec{z}} \right|^{-\Delta/2} G_{\Delta}(\hat{p}; \hat{q})
\ee
of \eqref{bbp}.
The Fourier transform \eqref{intm} can be evaluated to recover the conformal primary wavefunctions \eqref{mcpw}.

\subsection{Massless particles}

The massless conformal primary wavefunctions can be obtained from \eqref{mcpw} by taking the $m \rightarrow 0$ limit (assuming ${\rm Re}(\Delta) > 1$). Using the expansion 
\be 
\begin{split}
K_{\Delta - 1}(x) &= x^{ - \Delta} \left( 2^{\Delta -2}\Gamma(\Delta - 1)x + \mathcal{O}(x^2) \right) \\
&+ x^{\Delta} \left(2^{-\Delta} \Gamma(1 - \Delta) x^{-1} + 2^{-2 - \Delta} \frac{\Gamma(1 - \Delta)}{\Delta} x  + \mathcal{O}(x^2)\right),
\end{split}
\ee
we find
\be 
\label{msl-cpw}
\varphi_{\Delta}(\hat{q};x) = \lim_{m \rightarrow 0} \Psi_{\Delta}(\hat{q};x) \propto \frac{1}{(\hat{q} \cdot x)^{\Delta}}.
\ee
The integral representation of the massless conformal primary wavefunctions can also be derived from \eqref{intm} by taking the limit $m \rightarrow 0$ for fixed $\omega \equiv m/(2y)$. In this limit \eqref{massive-mom} becomes null, \eqref{bbp} becomes proportional to \cite{Kutasov:1999xu}
\be 
\label{btbpl}
\lim_{y \rightarrow 0} G_{\Delta}(y,z,\bz; w, \bw) = \frac{\pi}{\Delta - 1} y^{2 - \Delta} \delta^{(2)}(z - w) + \frac{y^{\Delta}}{(w - z)^{2\Delta}} + \mathcal{O}(y^{4 - \Delta}) 
\ee
and upon evaluating the integral in \eqref{intm} we recover \eqref{msl-cpw}.\footnote{In fact \eqref{btbpl} gives two contributions in the massless limit: the conformal primary \eqref{msl-cpw} and its shadow. Since these solutions are not linearly independent, \cite{Pasterski:2017kqt} argued it is sufficient to restrict to \eqref{msl-cpw}.}
\begin{exercise}
Evaluate \eqref{intm} in the limit \eqref{btbpl} and show that the massless conformal wavefunction indeed reduces to \eqref{msl-cpw}.
\end{exercise}

\subsection{A conformal primary basis for scattering in AFS}

Using the map \eqref{intm} from plane wave solutions to conformal primary solutions of the scalar wave equation, one can relate momentum space scattering amplitudes $\mathcal{A}$ to scattering amplitudes $\widetilde{\mathcal{A}}$ in a conformal primary basis 
\be 
\label{ca}
\widetilde{\mathcal{A}}(\Delta_i, z_i, \bz_i) = \prod_{i = 1}^n \int_{H_3} \frac{d^3\hat{p}_i}{p^0_i} G_{\Delta_i}(\hat{p}_i; z_i, \bz_i) \mathcal{A}(\epsilon_i m_i \hat{p}_i),
\ee
where $\epsilon_i = \pm 1$ depending on whether the $i^{\rm th}$ particle is incoming or outgoing. The transformation of \eqref{sl2c} under SL$(2, \mathbb{C})$ implies that \eqref{ca} transform as correlators of 2D (global) conformal primary operators 
\be 
\widetilde{\mathcal{A}}(\Delta_i, \vec{z}_i'(\vec{z}_i)) = \prod_{i = 1}^n \left|\frac{\p \vec{z}_i'}{\p \vec{z}_i} \right|^{-\Delta_i/2}\widetilde{\mathcal{A}}(\Delta_i, \vec{z}_i).
\ee
\eqref{ca} is the defining relation of a \textit{celestial amplitude}.

It can be shown that \cite{Pasterski:2017kqt}
\begin{itemize}
\item The massive conformal primary wavefunctions \eqref{intm} form a basis of solutions to the wave equation provided $\Delta = 1 + i\lambda,~ \lambda \geq 0$. For such $\Delta,$ these solutions are complete, linearly independent and orthogonal.\footnote{Notice that in this case, $p^2$ (the effective ``mass'' on the $AdS_3$ slices in \eqref{wes}) will be complex} 
\item The massless conformal primary wavefunctions form a basis of solutions to the wave equation provided $\Delta = 1 + i\lambda,~ \lambda \in \mathbb{R}.$
\end{itemize}
The construction of conformal primary wavefunctions and celestial amplitudes generalizes for spinning particles. Photons and gravitons are discussed in \cite{Pasterski:2017kqt, Donnay:2018neh}, fermions are analyzed in \cite{Narayanan:2020amh, Iacobacci:2020por} while arbitrary spins are addressed in \cite{Law:2020tsg, Pasterski:2020pdk}.

\subsection{Example: 2 massless and 1 massive scalars at tree--level}
\label{ex}

In this section we illustrate the construction of celestial amplitudes with a computation of the tree-level celestial amplitude for two massless and one massive scalars \cite{Lam:2017ofc}. We start with the momentum space 3-point interaction
\be 
\mathcal{A}(\hat{p}_i) = g \delta^{(4)}(\omega_1 \hat{q}_1 + \omega_2 \hat{q}_2 - m \hat{p}).
\ee
The associated celestial amplitude is then
\be 
\label{3ptca}
\begin{split}
\widetilde{\mathcal{A}}(\Delta_i, z_i, \bz_i) &= g \prod_{i = 1}^2\left(\int_0^{\infty} d\omega_i \omega_i^{\Delta_i - 1} \right) \int_0^{\infty} \frac{dy}{y^3} \int d^2w \left(\frac{y}{y^2  + |z_3 - w|^2} \right)^{\Delta_3}\\
&\times \delta^{(4)}(\omega_1 \hat{q}_1 + \omega_2 \hat{q}_2 - m \hat{p}).
\end{split}
\ee
Using the parameterizations of momenta
\be 
\begin{split}
\hat{q} &= \left(1 + z\bz, z + \bz, -i(z - \bz), 1 - z\bz \right), \\
\hat{p} &= \frac{1}{2y}\left(1 + y^2 + w\bw, w + \bw, -i(w - \bw), 1 - y^2 - w\bw \right)
\end{split}
\ee
and evaluating the integrals over $y, \vec{w}$ and $\omega_2$, \eqref{3ptca} reduces to (see appendix \ref{example} for details)
\be 
\label{int-res}
\widetilde{\mathcal{A}}(\Delta_i, z_i, \bz_i) = \frac{g m^{2\Delta_2 + \Delta_3 - 4}}{2^{2\Delta_2 - \Delta_3 - 2}|z_{12}|^{2\Delta_2 - 2\Delta_3}}\int_0^{\infty} d\omega \frac{\omega^{\Delta_1 - \Delta_2 + \Delta_3 - 1}}{\left(m^2|z_{23}|^2 + 4|z_{12}|^2|z_{13}|^2 \omega^2 \right)^{\Delta_3}}.
\ee
Upon a change of variables, the remaining integral becomes proportional to the standard integral 
\be 
\int_0^1 dt t^{\alpha - 1}(1 - t)^{\beta - 1} = \frac{\Gamma(\alpha) \Gamma(\beta)}{\Gamma(\alpha + \beta)}\equiv B(\alpha, \beta)
\ee
and we conclude 
\be 
\label{3-point}
\begin{split}
\widetilde{\mathcal{A}}(\Delta_i, z_i, \bz_i) &= \frac{C(\Delta_1, \Delta_2, \Delta_3)}{|z_{12}|^{\Delta_1 + \Delta_2 - \Delta_3}|z_{13}|^{\Delta_1 + \Delta_3 - \Delta_2}|z_{23}|^{\Delta_2 + \Delta_3 - \Delta_1}}, \\
C(\Delta_1, \Delta_2, \Delta_3) &= g \frac{m^{\Delta_1 + \Delta_2 - 4}}{2^{\Delta_1 + \Delta_2-1}} B\left(\frac{\Delta_{12} + \Delta_3}{2}, \frac{-\Delta_{12} + \Delta_3}{2} \right).
\end{split}
\ee
\eqref{3-point} takes the form of a 2D conformal 3-point correlator. In conformal field theory this is fixed up to a three-point coefficient by conformal symmetry.  We will see in the next section that in this case, this three-point coefficient is determined (up to a constant) by additionally imposing momentum conservation. Celestial three-point amplitudes of massless particles are more subtle as they involve singular conformally covariant three-point structures \cite{Pasterski:2017ylz}. It would be interesting to find a framework\footnote{Perhaps this may be achieved in a conformal primary basis involving light- or shadow-transforms of the massless conformal primaries \eqref{msl-cpw}.} in which all three-point celestial amplitudes are proportional to \eqref{3-point}. 

\section{Celestial symmetries}
\label{csymm}

This section contains a review of celestial symmetries and the constraints they impose on celestial amplitudes. The ideas and calculations summarized in this section are detailed in \cite{Stieberger:2018onx, Law:2019glh, Law:2020tsg, Pate:2019mfs, Pate:2019lpp, Guevara:2021abz, Strominger:2021lvk}.

\subsection{Poincar\'e action on the celestial sphere}

We begin by discussing the Poincar\'e symmetry of celestial amplitudes. As shown in \cite{Stieberger:2018onx} the Lorentz generators act on operators  $\mathcal{O}_{h, \bh}(z, \bz)$ as
\be 
\label{lg}
\begin{split}
L_0 \mathcal{O}_{h, \bh} &= 2(z \p_z + h)\mathcal{O}_{h, \bh}, \quad L_-\mathcal{O}_{h, \bh} = \p_z\mathcal{O}_{h, \bh}, \quad L_{+}\mathcal{O}_{h, \bh} = (z^2 \p_z + 2z h)\mathcal{O}_{h, \bh}, \\
\bar{L}_0\mathcal{O}_{h, \bh} &= 2(\bz\p_{\bz} + \bh)\mathcal{O}_{h, \bh}, \quad \bar{L}_-\mathcal{O}_{h, \bh} = \p_{\bz}\mathcal{O}_{h, \bh}, \quad \bar{L}_+\mathcal{O}_{h, \bh} = (\bz^2 \p_{\bz} + 2\bz \bh)\mathcal{O}_{h, \bh}.
\end{split}
\ee
%{\color{red} make sure the notation is consistent}
Lorentz symmetry of scattering in AFS is equivalent to global conformal symmetry of celestial amplitudes
\be 
\label{lc}
\mathcal{L}_I \widetilde{\mathcal{A}}_n = \bar{\mathcal{L}}_I \widetilde{\mathcal{A}}_n = 0,
\ee
where $\widetilde{\mathcal{A}}_n$ is an n-point celestial amplitude,
\be 
\mathcal{L}_I = \sum_{k = 1}^n L_{I, k}, \quad \bar{\mathcal{L}}_I \equiv \sum_{k = 1}^n \bar{L}_{I, k}
\ee
and $I$  runs over $-1, 0, 1$. \eqref{lc} is a familiar property of correlation functions in $2$D CFT.

Additionally, bulk translation invariance implies that
\be 
\label{mc}
\mathcal{P}_{\mu} \widetilde{\mathcal{A}}_n = 0, \quad \mathcal{P}_{\mu} \equiv \sum_{k = 1}^n P_{\mu,k}.
\ee
Celestial translation generators act on massless particles as weight-shifting operators
\be 
\label{mg}
P_{\mu, k} = \epsilon_k \hat{q}_{\mu}(z_k, \bz_k) e^{\p_{\Delta_k}},
\ee
where $\epsilon_k = \pm 1$ distinguishes between incoming and outgoing particles.
To see this, we can start with the momentum space action
\be 
\hat{P}_k A(q_1, ..., q_n) = \epsilon_k \omega_k \hat{q}_k A(q_1,..., q_n)
\ee
and rewrite it in a conformal primary basis by taking a Mellin transform
\be 
\begin{split}
P_k \widetilde{\mathcal{A}}(\Delta_1, ... \Delta_n) &= \prod_{j = 1}^n \left(\int_0^{\infty} d\omega_j \omega_j^{\Delta_j - 1} \right)\epsilon_k \omega_k \hat{q}_k A(q_1, ..., q_n) \\
&= \prod_{\substack{j = 1 \\ j \neq k}}^n \left(\int_0^{\infty} d\omega_j \omega_j^{\Delta_j - 1} \right) \int_0^{\infty} d\omega_k \omega_k^{\Delta_k +1 - 1} \epsilon_k \hat{q}_k A(q_1, ..., q_n)\\
& = \epsilon_k \hat{q}_k \widetilde{\mathcal{A}}(\Delta_1, ..., \Delta_k + 1, ..., \Delta_n).
\end{split}
\ee
We conclude that for massless scattering, \eqref{mc} relates celestial amplitudes involving operators of shifted weights
\be 
\widetilde{\mathcal{A}}_n(\Delta_1 + 1, \Delta_2, ..., \Delta_n) + \widetilde{\mathcal{A}}_n(\Delta_1, \Delta_2 + 1, ..., \Delta_n) + \cdots +  \widetilde{\mathcal{A}}_n(\Delta_1,..., \Delta_n + 1) = 0.
\ee

For massive scalars, \eqref{mg} are replaced by \cite{Law:2019glh}
\be 
\label{mmg}
P^{\mu} = \frac{m}{2}\left[\left(\p_z\p_{\bz} \hat{q}^{\mu} + \frac{\p_{\bz}\hat{q}^{\mu}\p_z + \p_z \hat{q}^{\mu}\p_{\bz}}{\Delta - 1} + \frac{\hat{q}^{\mu} \p_z\p_{\bz}}{(\Delta - 1)^2} \right) e^{-\p_{\Delta}} + \frac{\Delta \hat{q}^{\mu}}{\Delta - 1} e^{\p_{\Delta}} \right].
\ee
\eqref{mmg} is determined by imposing the on-shell condition
\be
\label{on-shell}
P_{\mu} P^{\mu} = -m^2, 
\ee
as well as the Poincar\'e algebra
\be 
\label{pa}
[P_{\mu}, P_{\nu}] = 0, \quad [M_{\mu\nu}, P_{\rho}] = \eta_{\mu\rho} P_{\nu} - \eta_{\nu \rho} P_{\mu}.
\ee
\begin{exercise}
Verify \eqref{mmg} satisfy \eqref{on-shell} and \eqref{pa}.
\end{exercise}

The momentum generators for spinning particles can be found in \cite{Law:2020tsg}. It is interesting to notice that in addition to off-diagonal terms in dimension $\Delta = h + \bh$, they also contain off-diagonal terms in spin $J = h - \bh$. 
%{\color{red} Intuition?}

\subsection{Constraints from Poincar\'e symmetry}

\eqref{lg} and \eqref{mg} imply that any celestial 4-point function can be put into the form 
\be 
\widetilde{\mathcal{A}}_4 = K_{h_i, \bh_i}(z_i, \bz_i) \delta(z - \bz) f^{h_i, \bh_i}(z, \bz),
\ee
where 
\be 
\label{ccov-f}
K_{h_i, \bh_i}(z_i, \bz_i) = \prod_{i < j = 1}^4 z_{ij}^{h/3 - h_i - h_j}\bz_{ij}^{\bh/3 - \bh_i - \bh_j}, \quad h = \sum_{i = 1}^4 h_i, ~~ \bh = \sum_{i = 1}^4 \bh_i
\ee
and 
\be 
\label{cr}
z = \frac{z_{13} z_{24}}{z_{12} z_{34}}, \quad \bz = \frac{\bz_{13} \bz_{24}}{\bz_{12} \bz_{34}}
\ee
are the $2$D conformally invariant cross-ratios. For massless scattering, \eqref{cr} are related to the bulk variables (Mandelstam invariants) by 
\be 
\label{zcr}
z = -\frac{t}{s}, \quad s = -(p_1 + p_2)^2, \quad t = -(p_1 + p_3)^2.
\ee
%{\color{red} what about $z$ for massive scattering?} 
Additionally, in this case momentum conservation \eqref{mc} implies that 
\be 
\label{tic}
\sum_{j = 1}^4 K_{h_j + \frac{1}{2}, \bh_j + \frac{1}{2}}(z_i, \bz_i) f^{h_j + \frac{1}{2}, \bh_j + \frac{1}{2}}(z, \bz) = 0.
\ee
\begin{exercise}
Show that 
\be 
\sum_{j = 1}^4 K_{h_j + \frac{1}{2}, \bh_j + \frac{1}{2}}(z_i, \bz_i) = 0.
\ee
\end{exercise}
\noindent Since the conformally covariant factor \eqref{ccov-f} is translationally invariant by itself, \eqref{tic} can be non-trivially obeyed if 
\be 
\label{tc}
f^{h_i + \frac{1}{2}, \bh_i + \frac{1}{2}}(z, \bz) = f^{h_j + \frac{1}{2}, \bh_j + \frac{1}{2}}(z, \bz), \quad \forall i, j.
\ee
By induction it can be shown that \eqref{tc} implies that 
\be 
\label{sol}
f^{h_i, \bh_i}(z, \bz) = f^{\beta, J_i}(z, \bz), \quad \beta = \sum_{i = 1}^4 (h_i + \bh_i) = \sum_{i = 1}^4 \Delta_i.
\ee 
It follows that celestial 4-point functions of scalars are parameterized by two variables: $\beta$ and $z$. $\beta$ is the variable dual to the center of mass energy under the Mellin transform, while $z$ is related to the bulk scattering angle via \eqref{zcr}.

More generally, \eqref{mc} will be obeyed by celestial amplitudes with both massive and massless particles, provided $P_{\mu, k}$ are chosen appropriately. For example, consider the three-point amplitude of two massless and one massive scalars computed in section \ref{ex}. The same result can be perhaps more easily recovered by considering the constraint
\be 
\label{mc3pt}
\left( P_1 + P_2 + P_3^{(m)} \right) \widetilde{\mathcal{A}}_3(1, 2, 3^{(m)}) =0,
\ee
where $P_1, P_2$ are the massless momentum generators in \eqref{mg} while $P_3^{(m)}$ is the massive momentum \eqref{mmg}. 
Global conformal invariance fixes 
\be 
\widetilde{\mathcal{A}}(1, 2, 3^{(m)}) = \frac{C(\Delta_1, \Delta_2, \Delta_3)}{|z_{12}|^{\Delta_1 + \Delta_2 - \Delta_3} |z_{23}|^{\Delta_2 + \Delta_3 - \Delta_1} |z_{13}|^{\Delta_1 + \Delta_3 - \Delta_2}},
\ee
hence \eqref{mc3pt} leads to the following recursion relations on the 3-point coefficients \cite{Law:2019glh}
\be 
\label{3-pt-c}
\begin{split}
\left(\frac{\Delta_{12}^2}{4} - \frac{(\Delta_3 - 1)^2}{4} \right) C_{\Delta_1, \Delta_2, \Delta_3 - 1} + \Delta_3(\Delta_3 - 1) C_{\Delta_1, \Delta_2, \Delta_3 + 1} = 0, \\
4\epsilon_2 (\Delta_3-1) C_{\Delta_1, \Delta_2 + 1, \Delta_3} + m \epsilon_3(\Delta_3 - 1 - \Delta_{12})C_{\Delta_1, \Delta_2, \Delta_3 - 1} = 0, \\
4\epsilon_1 (\Delta_3-1) C_{\Delta_1 + 1, \Delta_2, \Delta_3} + m \epsilon_3(\Delta_3 - 1 + \Delta_{12})C_{\Delta_1, \Delta_2, \Delta_3 - 1} = 0,
\end{split}
\ee
where $\Delta_{12} \equiv \Delta_1 - \Delta_2.$
\begin{exercise}
Show that \eqref{3-pt-c} are solved by 
\be 
C(\Delta_1, \Delta_2, \Delta_3) = B\left(\frac{\Delta_1 - \Delta_2 + \Delta_3}{2}, \frac{\Delta_2 - \Delta_1 + \Delta_3}{2} \right) c_{\Delta_1, \Delta_2, \Delta_3},
\ee
where 
\be 
\label{pc}
\begin{split}
c_{\Delta_1, \Delta_2, \Delta_3 - 1} = c_{\Delta_1, \Delta_2, \Delta_3 + 1}, \\
c_{\Delta_1 + 1, \Delta_2, \Delta_3} = c_{\Delta_1, \Delta_2 + 1, \Delta_3}, \\
c_{\Delta_1 + 1, \Delta_2, \Delta_3} = c_{\Delta_1, \Delta_2, \Delta_3 - 1}.
\end{split}
\ee
\end{exercise}
The first constraint in \eqref{pc} implies c is perodic in $\Delta_3$ of period 2, the second implies periodicity of period 1 in $\Delta_1$ and $ \Delta_2$ up to dependence of $\Delta_1 + \Delta_2$ while additionally, the last constraint implies $c$ is periodic of period 1 in $\Delta_1, \Delta_3$ up to dependence of $\sum_{i = 1}^3 \Delta_i$ in which case the period becomes 2. The periodic function is set to a constant by requiring the inverse Mellin transform to be well defined. 

\subsection{Conformally soft symmetries}
\label{sec:css}

We saw in section \ref{Virasoro} that the subleading soft theorem implies a conservation law associated with an infinite-dimensional Virasoro symmetry. This is one example of a more general equivalence between soft theorems and conservation laws associated with large gauge symmetries. For example, it was shown in \cite{He:2014cra} that the leading soft photon theorem implies that soft photons behave as $U(1)$ currents. As such, their insertions into $\mathcal{S}$-matrices obey Ward identities of the form
\be 
\label{u1c}
\begin{split}
\langle J_z \mathcal{O}_1(\omega_1, z_1, \bz_1)...\mathcal{O}_n(\omega_n, z_n, \bz_n)\rangle &\equiv \lim_{\omega \rightarrow 0}\omega \langle \mathcal{O}^+(\omega, z, \bz) \mathcal{O}_1(\omega_1, z_1, \bz_1)... \mathcal{O}_n(\omega_n, z_n, \bz_n)\rangle\\
&= \sum_{k = 1}^n \frac{Q_k}{z - z_k} \langle \mathcal{O}_1(\omega_1, z_1, \bz_1)...\mathcal{O}_n(\omega_n, z_n, \bz_n) \rangle.
\end{split}
\ee
Similarly, the leading soft gluon theorem can be recast as a holomorphic Kac-Moody symmetry generated by non-abelian currents $J_z^a$ (soft gluons of positive helicity) obeying the Ward identities \cite{He:2015zea}
\be 
\label{nac}
\begin{split}
\langle J_z^a \mathcal{O}_1(\omega_1, z_1, \bz_1)...\mathcal{O}_n(\omega_n, z_n, \bz_n)\rangle &\equiv \lim_{\omega \rightarrow 0} \omega \langle \mathcal{O}^{+, a}(\omega,z, \bz) \mathcal{O}_1(\omega_1, z_1, \bz_1)... \mathcal{O}_n(\omega_n, z_n, \bz_n) \rangle\\
&= \sum_{k = 1}^n \frac{1}{z - z_k} \langle \mathcal{O}_1(\omega_1, z_1, \bz_1)...T_k^a \mathcal{O}_k...\mathcal{O}_n(\omega_n, z_n, \bz_n) \rangle.
\end{split}
\ee

We would like to reexpress  \eqref{u1c} and \eqref{nac} in a conformal primary basis and identify the celestial representations of these symmetry generators. Since the currents were constructed from low-energy limits of bulk photons and gluons, while celestial operators involve integrals over photons and gluons of all energies, it is a-priori not immediately obvious how to construct the celestial currents. One hint is that in conventional CFT$_d$, currents saturate unitarity bounds\footnote{No analog bounds are known to exist in CCFT. Moreover, as we will see there is an infinite tower of negative dimension operators arising from soft limits in the bulk.} and hence the dimension of a spin-$j$ current is constrained to be \cite{Qualls:2015qjb}
\be 
\Delta = d + j - 2.
\ee
In particular, positive-helicity conformally soft photons and gluons should correspond to operators of weights\footnote{The conformal weights are related to the conformal dimensions $\Delta$ and the spin $J$ by $h = \frac{\Delta + J}{2},~ \bar{h} = \frac{\Delta - J}{2}$.} $(h, \bh) = (1, 0)$, while negative helicity ones should have $(h, \bh) = (0,1).$ They should be associated with abelian and non-abelian symmetries on the celestial sphere. 

The simplest way to show that this guess is indeed correct is to start with the Mellin representation 
\be 
\mathcal{O}_{\Delta}^+(z, \bz) = \int_0^{\infty}d\omega \omega^{\Delta - 1} \mathcal{O}^+(\omega, z, \bz)
\ee
and notice that  \cite{Pate:2019mfs}
\be 
\begin{split}
\lim_{\Delta \rightarrow 1}(\Delta - 1) \mathcal{O}_{\Delta}^+(z, \bz) &= \lim_{\Delta \rightarrow 1} \int_0^{\infty} d\omega (\Delta - 1) \omega^{\Delta - 1} \mathcal{O}^+(\omega, z, \bz) \\
&= 2\int_0^{\infty} d\omega \delta(\omega) \omega \mathcal{O}^+(\omega, z, \bz) = \lim_{\omega \rightarrow 0} \omega \mathcal{O}^+(\omega, z, \bz). 
\end{split}
\ee
In the last line we have used the identity\footnote{This holds provided that $x$ has compact support.}
\be 
\lim_{\epsilon \rightarrow 0} \frac{\epsilon}{2} |x|^{\epsilon - 1} = \delta(x).
\ee

More generally
\be 
\begin{split}
\lim_{\Delta \rightarrow -n} (\Delta + n) \mathcal{O}^+_{\Delta}(z, \bz) &= \lim_{\Delta \rightarrow -n} (\Delta + n) \int_0^{\omega_*} d\omega \omega^{\Delta - 1} \mathcal{O}^+(\omega, z, \bz)\\
&= \lim_{\Delta \rightarrow -n}(\Delta + n) \sum_{k} \int_0^{\omega_*}d\omega \omega^{\Delta + k - 1} O^+_k(z, \bz)\\
&= O^+_n(z, \bz),
\end{split}
\ee
where we expanded
\be 
\mathcal{O}^+(\omega, z, \bz) = \sum_k \omega^k O^+_k(z, \bz)
\ee
for $\omega \ll \omega_*$ and assumed that insertions of $\mathcal{O}^+(\omega, z, \bz)$ into $\mathcal{S}$-matrices have fast enough fall-offs with energy\footnote{An exponential fall-off $\lim_{\omega \rightarrow \infty}\langle \mathcal{O}(\omega, z, \bz)\cdots \rangle \sim e^{-\epsilon \omega}$ will ensure this limit is well defined for any negative integer $\Delta$.} in which case the  high-energy part of the Mellin integral will be free of poles in $\Delta + n$. We conclude that the $\Delta \rightarrow -n$ limit of a celestial operator for $n = -1, 0, 1,...$ picks out the $\mathcal{O}(\omega^n)$ term in an expansion around $\omega = 0$. For example, a subleading soft photon will correspond to the celestial operator
\be
\lim_{\Delta \rightarrow 0} \Delta \mathcal{O}^+_{\Delta}(z, \bz).
\ee
This infinity of soft currents has been studied in \cite{Guevara:2019ypd, Adamo:2019ipt,Guevara:2021abz,Strominger:2021lvk}. There exists a complementary tower of positive integer-dimension operators (also known as Goldstone modes), canonically conjugate to the conformally soft modes above \cite{Donnay:2020guq}. Their combined Ward identities are expected to constrain celestial amplitudes, but a complete understanding of these symmetries and their implications remains an important open problem. In the next section we describe some instances in which soft celestial symmetries were used to  derive non-trivial properties of celestial amplitudes.

\subsection{Applications}

We conclude with brief overview of recent work on soft constraints on celestial amplitudes \cite{Pate:2019lpp} and the infinite tower of soft currents \cite{Guevara:2021abz, Strominger:2021lvk}. 

\subsubsection{Celestial operator products of gluons}
We start  by assuming that positive-helicity gluons admit the holomorphic collinear expansion
\be 
\label{gOPE}
\mathcal{O}^{+, a}_{\Delta_1}(z_1, \bz_1) \mathcal{O}^{+, b}_{\Delta_2}(z_2, \bz_2) \sim -\frac{i f^{ab}_{~~c}}{z_{12}} C(\Delta_1, \Delta_2) \mathcal{O}^{+, c}_{\Delta_1 + \Delta_2 - 1}(z_2, \bz_2) + \cdots,
\ee
where $...$ include contributions from SL$(2, \mathbb{C})$ descendants. Here and in the next section, $z, \bz$ are treated as real independent variables in which case the CCFT becomes Lorentzian and SL$(2, \mathbb{C})$ is replaced by SL$(2,\mathbb{R})_{\rm L} \times$ SL$(2, \mathbb{R})_{\rm R}$. The form of the OPE is fixed by the leading soft theorem and SL$(2,\mathbb{C})$ up to a coefficient $C(\Delta_1, \Delta_2)$. We now show that the subleading conformally soft gluon theorem determines this leading OPE coefficient up to a normalization fixed by the leading soft gluon theorem \cite{Pate:2019lpp}.

The subleading soft gluon theorem can be recast as a ``symmetry''\footnote{These have not been shown to be associated with asymptotic charges.} under which gluons transform as follows 
\be 
\begin{split}
\delta_b \mathcal{O}^{\pm, a}_{\Delta}(z, \bz) = -(\Delta - 1 \pm 1 + z\p_z) i f^a_{\ bc} \mathcal{O}^{\pm, c}_{\Delta - 1}(z,\bz), \\
\bar{\delta}_b \mathcal{O}^{\pm, a}_{\Delta}(z, \bz) = -(\Delta - 1 \mp 1 + \bz\p_{\bz}) i f^a_{\ bc} \mathcal{O}^{\pm, c}_{\Delta - 1}(z,\bz).
\end{split}
\ee
Acting with $\bar{\delta}$ on both sides of \eqref{gOPE} and comparing the two sides, we deduce that $C(\Delta_1, \Delta_2)$ obey the recursion relation
\be 
\label{rr}
(\Delta_1 - 2) C(\Delta_1 - 1, \Delta_2) = (\Delta_1 + \Delta_2 - 3) C(\Delta_1, \Delta_2).
\ee
\eqref{rr} has the unique\footnote{By Wieland's theorem, see appendix E of \cite{Pate:2019lpp}. The normalization is fixed by the leading soft theorem.} solution
\be 
C(\Delta_1, \Delta_2) = B(\Delta_1 - 1, \Delta_2 - 1).
\ee
\begin{exercise}
Write down the action of $\bar{\delta}$ on \eqref{gOPE} and deduce \eqref{rr}.
\end{exercise}
Similar recursion relations are also implied by the subsubleading soft graviton theorem and can be shown to completely fix the leading OPE coefficients in Einstein-Yang-Mills theory. 

\subsubsection{Holographic symmetry algebras}

\eqref{gOPE} can be generalized to include contributions from SL$(2, \mathbb{R})_R$ descendants. One finds
\be 
\label{Oped}
\mathcal{O}^{+,a}_{\Delta_1}(z_1, \bz_1) \mathcal{O}_{\Delta_2}^{+, b}(z_2, \bz_2) \sim \frac{-i f^{ab}_{~~ c}}{z_{12}} \sum_{n = 0}^{\infty} B(\Delta_1 - 1 + n, \Delta_2 - 1) \frac{\bz_{12}^n}{n!} \bar{\p}^n\mathcal{O}^{+,c}_{\Delta_1 + \Delta_2 - 1}(z_2, \bz_2).
\ee
\eqref{Oped} follows by resuming contributions from the right-moving descendants through the OPE block \cite{Guevara:2021abz}
\be 
\mathcal{O}^{+,a}_{\Delta_1}(z_1, \bz_1) \mathcal{O}_{\Delta_2}^{+, b}(z_2, \bz_2) \sim \frac{-i f^{ab}_{~~ c}}{z_{12}} \int_0^1 dt \frac{\mathcal{O}_{\Delta_P}^{+,c}(z_2, \bz_2 + t \bz_{12})}{t^{2 - \Delta_1}(1 - t)^{2 - \Delta_2}}.
\ee
It is interesting to study the algebra of soft operators discussed in section \ref{sec:css}. First notice that if $\Delta_1, \Delta_2 \in \{1, 0, -1,...\}$, $\Delta_1 + \Delta_2 - 1 \in \{1, 0, -1,...\}$ and the algebra of soft operators closes. Then mode expanding such an operator on the right, one finds for $k = 1, 0, -1,...$
\be 
\lim_{\epsilon \rightarrow 0} \epsilon \mathcal{O}^{+,a}_{k + \epsilon}(z, \bz) =\lim_{\epsilon \rightarrow 0} \sum_{n} \frac{\epsilon \mathcal{O}^{+,a}_{k + \epsilon, n}(z)}{\bz^{n + \frac{k - 1}{2}}}.
\ee
Defining
\be 
\label{scurr}
R^{k, a}(z, \bz) = \lim_{\epsilon \rightarrow 0} \epsilon \mathcal{O}^{+,a}_{k + \epsilon}(z, \bz), \quad R^{k, a}_n(z) = \lim_{\epsilon \rightarrow 0} \epsilon \mathcal{O}_{k + \epsilon,n}^{+,a}(z), 
\ee
we see that for 
\be 
\frac{k  -1}{2} \leq n \leq \frac{1 - k}{2},
\ee
$R^{k,a}_n(z)$ organize into $(2 - k)$-dimensional SL$(2, \mathbb{R})_R$ representations as 
\be 
\bar{\p}^{2 - k} R^{k, a}(z, \bz) = 0.
\ee
Using \eqref{Oped} and \eqref{scurr}, the OPE of soft currents is found to be 
\be 
\label{soft-algebra}
R^{k, a}(z_1, \bz_1) R^{l, b}(z_2, \bz_2) \sim \frac{-i f^{ab}_{~~ c}}{z_{12}} \sum_{n = 0}^{1 - k} \left(\begin{matrix}
2 - k - l - n\\
1 - l
\end{matrix} \right) \frac{\bz_{12}^n}{n!} \bar{\p}^n R^{k + l - 1, c}(z_2, \bz_2).
\ee
This follows from setting $\Delta_1 = k + \epsilon, \Delta_2 = l + \epsilon$ in \eqref{Oped} and using
\be 
\lim_{\epsilon \rightarrow 0} \epsilon \frac{\Gamma(k + \epsilon - 1 +n)\Gamma(l + \epsilon - 1)}{\Gamma(k + l + 2\epsilon + n - 2)} = \frac{1}{(1 - l)!} \frac{\Gamma(3 - k - l - n)}{\Gamma(2 - k - n)}.
\ee
Finally, \eqref{soft-algebra} can be used to derive the algebra of the soft operators \cite{Guevara:2021abz}
\be 
\label{saf}
[R^{k, a}_n, R^{l, b}_{n'}] = -i f^{ab}_{~~ c}\left(\begin{matrix}
\frac{1 - k}{2} - n + \frac{1 - l}{2} - n'\\
\frac{1 - k}{2} - n
\end{matrix} \right) \left(\begin{matrix} \frac{1 - k}{2} + n + \frac{1 - l}{2} + n'\\
\frac{1 - k}{2} + n\end{matrix}\right) R^{k + l - 1, c}_{n + n'},
\ee
which follows from
\be 
R_n^{k, a}(z) = \oint \frac{d\bz}{2\pi i} \bz^{n + \frac{k - 1}{2} - 1} R^{k, a}(z, \bz)
\ee
and the commutator of holomorphic operators 
\be 
[A, B](z) =\oint_z \frac{dw}{2\pi i} A(w) B(z)
\ee
applied to \eqref{soft-algebra}.

Notice that upon redefining \cite{Strominger:2021lvk}
\be 
\hat{R}^{k, a}_n \equiv \left(\frac{1 - k}{2} - n\right)! \left(\frac{1 - k}{2} + n\right)! R^{k, a}_n,
\ee
\eqref{saf} simplifies to 
\be 
\label{saf1}
[\hat{R}^{k, a}_n, \hat{R}^{l, b}_{n'}] = -i f^{ab}_{~~ c} \hat{R}^{k + l - 1, c}_{n + n'}.
\ee
A similar analysis can be done for gravitons, where the analog of \eqref{saf1} was identified with a $w_{1 + \infty}$ algebra in \cite{Strominger:2021lvk}
\be 
\label{w-alg}
[w^p_m, w^q_n] = [m(q - 1) - n(p - 1)] w_{m + n}^{p + q - 2},
\ee
with $p, q$ running over positive, half-integral values $p, q = 1, \frac{3}{2},\cdots$.
Working out the implications of \eqref{w-alg} for gravity in AFS remains a fascinating open problem. 

\section*{Acknowledgements}
I would first like to thank the organizers Alejandra Castro, Bruno Carneiro da Cunha, Thiago Fleury, Dmitry Melnikov and Jo$\tilde{\rm a}$o Penedones for the opportunity to speak at the Pre-Strings school 2021 and especially Alejandra for encouraging me to publish these notes. I would also like to thank the students and other lectures for many interesting discussions. I am particularly grateful to friends and collaborators including Alex Atanasov, Adam Ball, Laura Donnay, Laurent Freidel, Alfredo Guevara, Temple He, Mina Himwich, Dan Kapec, Walker Melton, Noah Miller, Prahar Mitra, Sruthi Narayanan, Sabrina Pasterski, Monica Pate, Andrea Puhm, Andy Strominger, Tomasz Taylor and Ellis Yuan from whom I have learned so much about the topics discussed here. I acknowledge support from the Stephen Hawking Postdoctoral Fellowship at Perimeter Institute. Research at Perimeter Institute is supported in part by the Government of Canada through the Department of Innovation, Science and Industry Canada and by the Province of Ontario through the Ministry of Colleges and Universities.

\appendix
\section{Charge commutators}
\label{commutation-rel}
In this section we compute the commutators of the charges \eqref{charges} with $C_{ww}$. Upon integration by parts the hard charges can be rewritten as
\be 
\begin{split}
Q^+_H = \frac{1}{4}\int_{\mathcal{I}^+}dud^2z \gamma_{z\bz}&\Big(u D_zY^{+z} N_{zz}N^{zz} + u D_{\bz} Y^{+\bz} N_{\bz\bz}N^{\bz\bz} 
+ D_zY^{+z} C_{zz}N^{zz} \\
&+ D_{\bz}Y^{+ \bz} C_{\bz\bz}N^{\bz\bz} 
+ 2D_z(Y^{+z} C_{zz})N^{zz} + 2Y^{+\bz} N_{\bz\bz} D_{\bz} C^{\bz\bz} \Big) + {\rm matter}.
\end{split}
\ee
Then using the canonical commutation relations
\be 
[N_{\bz\bz}(u, z, \bz), C_{ww}(u', w, \bw)] = 2i \gamma_{z\bz} \delta^{(2)}(z - w) \delta(u - u'),
\ee
one derives
\be 
\begin{split}
[Q_H^+, C_{ww}(u', w, \bw)] &= \frac{2i}{4}\Big(u D\cdot Y^+ N_{ww} + (D_wY^{+w} - D_{\bw} Y^{+\bw}) C_{ww} + 2D_w(Y^{+w} C_{ww}) \\
&+ 2Y^{+\bw} D_{\bw} C_{ww} \Big) = i \Big(\frac{u}{2} D\cdot Y^+ N_{ww} - \frac{1}{2}D\cdot Y^+ C_{ww}  \\
&+ \underbrace{2 D_w Y^{+w} C_{ww} + Y^{+w} D_w C_{ww}  +  Y^{+\bw} D_{\bw} C_{ww}}_{\mathcal{L}_{Y^+} C_{ww}} \Big) = i\delta_{Y^+}^H C_{ww}.
\end{split}
\ee
Similarly,
\be 
[Q_S^+, C_{ww}] = -i u D^3_w Y^{+w} = i\delta_{Y^+}^S C_{ww}.
\ee
The commutator of $Q^+$ with $N_{ww}$ is derived analogously.

\section{Conformal primaries in Milne coordinates}
\label{mwcp}

The relation between the Lorentz generators $J_i, K_i$ and the SL$(2, \mathbb{C})$ generators $L_i, \bar{L}_i$ is
\be 
\label{genn}
\begin{split}
L_0 &= -\frac{i}{2}(J_3 + i K_3),\quad\qquad \qquad \quad~~ \bar{L}_0 = \frac{i}{2}(J_3 - i K_3),\\
L_{1} &= -\frac{i}{2}(J_1 + iK_1 + i(J_2 + iK_2)),~~ \bar{L}_{1} = \frac{i}{2}(J_1 - iK_1 - i(J_2 - iK_2)),
\\ L_{-1} &= \frac{i}{2}(J_1 + iK_1 - i(J_2 + iK_2)), ~~~\bar{L}_{-1} = -\frac{i}{2}(J_1 - iK_1 + i(J_2 - iK_2)). \\
\end{split}
\ee
The Lorentz algebra \eqref{clg} immediately implies that \eqref{genn} obey the SL$(2,\mathbb{C})$ algebra \eqref{clgp}.

The form \eqref{H3} of the Minkowski metric with $\rho = \sinh \eta$ can be obtained directly from
\be 
\label{Minkn}
ds^2 = -(dx^0)^2 + (dx^1)^2 + (dx^2)^2 + (dx^3)^2
\ee
via the coordinate transform
\be 
\begin{split}
x^0 &= \tau \cosh \eta,\\
x^1 &= \tau \sin \theta \cos \varphi \sinh \eta,\\
x^2 &= \tau \sin \theta \sin  \varphi \sinh \eta,\\
x^3 &= \tau \cos \theta \sinh \eta.\\
\end{split}
\ee

The isometries of \eqref{H3} are inherited from isometries of \eqref{Minkn} which preserve the slices of constant $\tau$ and hence coincide with the Lorentz transformations \eqref{rot}, \eqref{boosts}.
In $(\tau, \eta, \theta, \varphi)$ coordinates, the Lorentz generators take the form
\be 
\label{LGsl}
\begin{split}
&J_3 =- \p_{\varphi},~~ J_1 = \sin \varphi \p_{\theta} + \cos \varphi \cot \theta\p_{\varphi},~~ J_2 =  -\cos \varphi \p_{\theta} + \sin \varphi \cot \theta\p_{\varphi},\\
&K_3 = -(\cos \theta\p_{\eta} -\sin \theta \coth \eta  \p_{\theta}),\\
&K_1 = -(\cos \varphi \sin \theta \p_{\eta}  + \cos \theta \cos \varphi \coth \eta \p_{\theta} -\coth \eta \csc \theta \sin \varphi \p_{\varphi}),\\
& K_2 = -(\sin \theta \sin \varphi \p_{\eta} +\cos\theta \coth \eta \sin \varphi \p_{\theta} + \coth \eta \cos \varphi \csc \theta \p_{\varphi}). ~~
\end{split}
\ee
In the limit $\eta \rightarrow \infty$, \eqref{LGsl} reduce to 
\be 
\label{pull-backs-lg}
\begin{split}
&J_3 = -\p_{\varphi},~~ J_1 = \sin \varphi \p_{\theta} + \cos \varphi \cot \theta\p_{\varphi},~~ J_2 = -\cos \varphi \p_{\theta} + \sin \varphi \cot \theta\p_{\varphi},\\
&K_3 =  \sin \theta  \p_{\theta},~~ K_1 = -( \cos \theta \cos \varphi  \p_{\theta} - \csc \theta \sin \varphi \p_{\varphi}),~~  K_2 = -(\cos\theta \sin \varphi \p_{\theta} + \cos \varphi \csc \theta \p_{\varphi}).
\end{split}
\ee
Equivalently in $(z, \bz)$ coordinates\footnote{$z \rightarrow -z, \bz \rightarrow -\bz$ is an automorphism of the Lorentz algebra. \eqref{stereo} yield formulas for the Lorentz generators that match with \cite{Kapec:2014opa} up to an overall sign. This overall sign is such that the standard Lorentz algebra \eqref{clg} is obeyed.}
\begin{equation}
\label{stereo}
z = -\cot \frac{\theta}{2} e^{i\varphi}, \qquad \bz = -\cot \frac{\theta}{2} e^{-i \varphi}
\end{equation}
and using identities such as
\be 
\frac{1}{(\sin \theta/2)^2} = 1 + z\bz,
\ee
\eqref{pull-backs-lg} take the form
\be 
\begin{split}
J_3 &= -i(z\p_z  - \bz\p_{\bz}),\qquad\qquad\qquad\qquad~~
K_3 = -(z\p_z + \bz\p_{\bz}) ,\\ J_1 &= -\frac{i}{2}\left[(z^2 - 1)\p_z - (\bz^2 - 1) \p_{\bz}\right],\qquad ~
J_2 = -\frac{1}{2}\left[(z^2 + 1)\p_z + (\bz^2 + 1)\p_{\bz}\right],\\
K_1 &= -\frac{1}{2}\left[(z^2 - 1)\p_z + (\bz^2 - 1) \p_{\bz}\right]  ,\qquad K_2 = \frac{i}{2}\left[(z^2 + 1)\p_z - (\bz^2 + 1)\p_{\bz}\right].  ~~
\end{split}
\ee
These precisely agree with \eqref{vf} with $f = 0$ and $Y^{+z}$ given in \eqref{gkvf}. 

As before,
\be 
\label{cps}
\Psi_{\Delta} = \frac{f(\tau^2)}{(x^0 + x^3)^{\Delta}} = \frac{f(\tau^2)}{\left(\tau(\cosh\eta + \cos \theta  \sinh\eta)\right)^{\Delta}}
\ee 
obeys 
\be 
\label{ev}
(L_0 + \bar{L}_0)\Psi_{\Delta} = \Delta \Psi_{\Delta}, \quad (L_0 - \bar{L}_0) \Psi_{\Delta} = 0
\ee
and 
\be 
\label{hw}
L_1 \Psi_{\Delta} = \bar{L}_1\Psi_{\Delta} = 0.
\ee
\eqref{cps} diagonalizes boosts along the $x^3$ axis and obeys the highest weight condition \eqref{hw}.

\section{Celestial 3-point example}
\label{example}

In this appendix we spell out the steps involved in evaluating the integral \eqref{3ptca}. We first notice that on the support of the momentum-conserving delta function,
\be 
\begin{split}
\omega_2 &= \frac{m^2}{4\omega_1|z_{12}|^2}, ~~ y = \frac{2m \omega_1|z_{12}|^2}{m^2 + 4\omega_1^2|z_{12}|^2},\\
w &= \frac{m^2 z_2 + 4\omega_1^2 z_1 |z_{12}|^2}{m^2 + 4\omega_1^2 |z_{12}|^2}, ~~ \bw = \frac{m^2 \bz_2 + 4\omega_1^2 \bz_1 |z_{12}|^2}{m^2 + 4 \omega_1^2 |z_{12}|^2}.
\end{split}
\ee
The Jacobian for the transformation from $(\omega_i \hat{q}_i, m \hat{p})$ to $(\omega_2, y, w, \bw)$ is 
\be 
|J| = \frac{m^3(y^2 + |w - z_2|^2)}{2 y^4}.
\ee
We then find that the integrand of \eqref{3ptca} is proportional to
\be 
\begin{split}
&\frac{1}{y^3} \frac{1}{|J|} \left(\frac{y}{y^2 + |w - z_3|^2} \right)^{\Delta_3} \delta\left(\omega_2 - \frac{m^2}{4\omega_1|z_{12}|^2}\right)\delta\left(y -\frac{2m \omega_1|z_{12}|^2}{m^2 + 4\omega_1^2|z_{12}|^2}\right)\\
&\times\delta^{(2)}\left(w - \frac{m^2 z_2 + 4\omega_1^2 z_1 |z_{12}|^2}{m^2 + 4\omega_1^2 |z_{12}|^2}\right) = \frac{2}{m^3} \frac{m}{2\omega_1|z_{12}|^2}\left(\frac{2m \omega_1 |z_{12}|^2}{4\omega_1^2|z_{12}|^2|z_{13}|^2 + m^2|z_{23}|^2} \right)^{\Delta_3}\times \delta^{(4)}.
\end{split}
\ee
The integrals over $y, w, \bw$ are now trivial and the celestial 3-point amplitude becomes
\be  
\begin{split}
\widetilde{\mathcal{A}}(\Delta_i, z_i, \bz_i) &= g \left(\frac{m^2}{4|z_{12}|^2} \right)^{\Delta_2 - 1} \frac{(2m|z_{12}|^2)^{\Delta_3}}{m^2 |z_{12}|^2}\int_0^{\infty}d\omega_1 \frac{\omega_1^{\Delta_1 - \Delta_2  + \Delta_3 - 1}}{(4\omega_1^2|z_{12}|^2|z_{13}|^2 + m^2|z_{23}|^2)^{\Delta_3}}\\
&= \frac{g m^{2\Delta_2 + \Delta_3 - 4}}{2^{2\Delta_2 - \Delta_3 - 2}|z_{12}|^{2\Delta_2 - 2\Delta_3}}\int_0^{\infty}d\omega_1 \frac{\omega_1^{\Delta_1 - \Delta_2  + \Delta_3 - 1}}{(4\omega_1^2|z_{12}|^2|z_{13}|^2 + m^2|z_{23}|^2)^{\Delta_3}},
\end{split}
\ee
which precisely agrees with \eqref{int-res}.

\bibliographystyle{utphys}
\bibliography{references}

\end{document}